\documentclass[aps,prx,twocolumn,floatfix,longbibliography,nofootinbib,superscriptaddress]{revtex4-1}
\usepackage[utf8]{inputenc}
\usepackage{natbib}
\usepackage{graphicx}
\usepackage{xcolor}
\usepackage[tbtags]{amsmath}
\usepackage[colorlinks, linkcolor=blue]{hyperref}
\hypersetup{colorlinks,allcolors=blue}
\usepackage{amssymb}
\usepackage{amsmath}
\usepackage{gensymb}
\usepackage{float}
\usepackage{comment}
\usepackage{amsmath}
\usepackage{tabularx,graphicx}
\usepackage{epstopdf}
\usepackage{latexsym}
\usepackage{color, colortbl}
\usepackage{psfrag}
\usepackage{bbm}
\usepackage{bm}
\usepackage{lmodern}
\usepackage{fix-cm}
\usepackage{xfrac}
\usepackage{color}    
\usepackage{lipsum}
\usepackage{textcomp}
\usepackage{amsmath}
\usepackage{mathtools}
\usepackage{subfigure}
\usepackage{titlesec}
\usepackage{dsfont}
\usepackage{feynmp}
\usepackage{slashed}
\usepackage{multirow}
\usepackage[normalem]{ulem}
\usepackage{svg}
\usepackage{braket}
\usepackage{nicematrix}

\newcommand{\bcen}{\begin{center}}
\newcommand{\ecen}{\end{center}}
\newcommand{\btab}{\begin{tabular}}
\newcommand{\etab}{\end{tabular}}
\newcommand{\bdes}{\begin{description}}
\newcommand{\edes}{\end{description}}

\newcommand{\beq}{\begin{equation}}
\newcommand{\eeq}{\end{equation}}
\newcommand{\bea}{\begin{eqnarray}}
\newcommand{\eea}{\end{eqnarray}}

\newcommand{\bary}{\begin{array}}
\newcommand{\eary}{\end{array}}
\newcommand{\benum}{\begin{enumerate}}
\newcommand{\eenum}{\end{enumerate}}
\newcommand{\bitem}{\begin{itemize}}
\newcommand{\eitem}{\end{itemize}}

\newcommand{\Fig}[1]{Fig.~\ref{#1}}

\makeatletter

\newcommand{\Rmnum}[1]{\expandafter\@slowromancap\romannumeral #1@}
\makeatother

\renewcommand{\textcolor}[2]{#2}    

\begin{document}

\title{Anyon Quasilocalization in a Quasicrystalline Toric Code}

\author{Soumya Sur}
\email{ssoumya@iitk.ac.in}
\affiliation{Department of Physics, Indian Institute of Technology Kanpur, Kalyanpur, UP 208016, India}

\author{Mohammad Saad}
\affiliation{Department of Physics, Indian Institute of Technology Kanpur, Kalyanpur, UP 208016, India} 
\affiliation{Department of Physics, University of Illinois Urbana-Champaign, Urbana, Illinois 61801, USA}

\author{Adhip Agarwala}
\email{adhip@iitk.ac.in}
\affiliation{Department of Physics, Indian Institute of Technology Kanpur, Kalyanpur, UP 208016, India}

\begin{abstract}
An exactly solvable model of a quantum spin liquid on a quasicrystal, akin to Kitaev's honeycomb model,  was introduced in Kim \textit{et al.}, \href{https://doi.org/10.1103/PhysRevB.110.214438}{\text{Phys. Rev. B} \textbf{110}, 214438 (2024)}. It was shown that in contrast to the translationally invariant models, such a spin liquid stabilizes a gapped ground state with a finite irrational flux density. In this work, we analyze the strong bond-anisotropic limit of the model and demonstrate that the aperiodic lattice geometry naturally generates a hierarchy of exponentially separated coupling constants in the resulting toric code Hamiltonian. Furthermore, a perturbative magnetic field leads to anomalous localization properties where an anyonic excitation sequentially delocalizes over subsets of sites forming equipotential contours in the quasicrystal. In addition, certain background flux configurations, together with the underlying geometry, give rise to strictly localized eigenstates that remain decoupled from the rest of the spectrum. Using numerical studies, we uncover the key mechanisms responsible for this unconventional localization behavior. Our study highlights that topologically ordered phases, in the presence of geometrical constraints can lead to highly anomalous localization properties of fractionalized charges.

\end{abstract}

\maketitle

\section{Introduction}
Quasicrystals, with their aperiodic yet long-range ordered lattice structure, remain one of the most enigmatic platforms to realize non-trivial phases of quantum matter. First realized in metallic alloys and magnetic materials \cite{PhysRevLett.53.1951, PhysRevLett.53.2477, RevModPhys.65.213, goldman1991quasicrystal, bindi2009natural}, a variety of exotic phases have been realized in them including non-fermi liquids \cite{deguchi2012quantum}, superconductivity \cite{kamiya2018discovery, uri2023superconductivity} and unconventional 
magnetism \cite{tamura2021experimental, tamura2025observation, chen2025quasicrystallinealtermagnetism}. Recently in twisted bilayer graphene, an emergent quasicrystal was realized \cite{pezzini202030, deng2020interlayer}. They have also been realized in non-solid-state platforms such as photonic lattices \cite{levi2011disorder, wang2024observation, lin2025photonic}, cold-atomic optical lattices \cite{PhysRevA.72.053607, PhysRevLett.122.110404, yu2024observing} and quantum computational setups \cite{lopez2023field}. With substantial experimental developments underway, theoretical studies have likewise expanded considerably.

Earlier work by Levine and others \cite{PhysRevB.34.596, PhysRevB.34.617, PhysRevB.32.5547, PhysRevB.37.8145, RevModPhys.63.699} has provided comprehensive classification and construction methods for quasicrystals. Being neither perfectly periodic nor completely disordered, quasicrystalline non-interacting tight-binding models exhibit rich spectral properties in both one and higher dimensions: they feature fully localized states as well as Cantor-like critical spectra that are neither localized nor extended \cite{PhysRevB.35.1020, PhysRevLett.56.2740, PhysRevB.38.1621}. These spectral features lead to unusual dynamical and transport properties in both non-interacting \cite{PhysRevB.46.13751, PhysRevB.62.15569, PhysRevB.99.224204} and interacting systems \cite{Vidal-Dominique-Giamarchi, lev2017transport, PhysRevB.100.104204, PhysRevB.106.184209, PhysRevB.106.094206}.

In the context of magnetism, various classical and quantum spin models have been studied on quasicrystalline lattices. It has been shown that quasiperiodic one-dimensional models belong to a critical universality class that is fundamentally distinct from both clean and disordered systems \cite{PhysRevLett.120.175702, agrawal2020universality}. In higher dimensions, finite antiferromagnetic order can emerge even when local electronic correlations are weak, owing to the presence of strictly localized states allowed by the quasicrystalline geometry \cite{PhysRevB.96.214402} and the resulting spatial inhomogeneity in the magnetization profile \cite{PhysRevLett.90.177205}. In the context of frustrated magnetism, studies on classical and quantum dimer models \cite{PhysRevX.10.011005, PhysRevB.106.094202}, some of which are exactly solvable \cite{PhysRevB.108.014426}, reveal nontrivial dimer ground states and characteristic dimer–monomer correlations that arise directly from the underlying nontrivial lattice symmetries.

While strongly correlated Landau-like phases in quasicrystals have been explored, the study of topological phases in such systems remains relatively sparse. Recent works have examined free-fermionic symmetry-protected topological phases \cite{PhysRevLett.109.106402, Hofstadter-Fuchs-vidal, PhysRevLett.123.196401, PhysRevLett.124.036803, PhysRevResearch.2.033071, PhysRevLett.129.056403, fan2022topological}, whereas the study of topologically ordered phases is still in its early stages \cite{PhysRevB.101.115413, PhysRevX.11.041051}. The latter, which are characterized by fractionalized excitations and long-range entanglement, are of central importance from both theoretical \cite{RevModPhys.89.041004, savary2016quantum, broholm2020quantum} and quantum computational perspectives \cite{semeghini2021probing, satzinger2021realizing}.

Significant progress has been made in classifying topologically ordered spin-liquid phases using projective symmetry group approaches on regular, translationally symmetric lattices \cite{PhysRevB.65.165113, PhysRevB.93.165113, PhysRevB.84.024420, PhysRevB.83.224413, PhysRevB.107.134438}. Realizing spin liquids on non-crystalline platforms opens new possibilities for discovering such phases and gaining new physical insights. For instance, a recent \textcolor{blue}{theoretical construction} of a Kitaev-like \cite{kitaev2006anyons} spin liquid on an amorphous lattice demonstrated that the ground-state flux spontaneously breaks time-reversal symmetry, giving rise to a chiral quantum spin liquid \cite{cassella2023exact, PhysRevLett.130.186702}. More recently, the concept of quantum spin liquids has been extended to non-Euclidean hyperbolic lattices \cite{s25y-s4fj, PhysRevLett.134.256604}, leading to exotic gapped and gapless chiral phases with non-Abelian Bloch excitations.

In this context, a recent study has shown that an exactly solvable model of a quantum spin liquid can be realized on a generalized Penrose-tiling quasicrystal \cite{PhysRevB.110.214438}. Unlike higher-spin generalizations \cite{PhysRevB.108.104208} and random amorphous setups \cite{cassella2023exact}, this work demonstrates that even in a spin-$\frac{1}{2}$ system, a gapped spin-liquid phase can emerge without breaking time-reversal symmetry. Moreover, its ground state hosts an irrational density of $\mathbb{Z}_2$ fluxes, stabilized by a generalized Lieb’s theorem \cite{PhysRevLett.73.2158} applicable to quasicrystalline lattices.

One of the primary reasons for the immense interest in Kitaev’s honeycomb model, beyond its exact solvability and potential experimental realizations \cite{trebst2022kitaev, kitaev-rydberg-prr, delfino2026ksl}, lies in its smooth connection to the toric code (TC) limit, which can be readily achieved by increasing the bond anisotropy \cite{kitaev2006anyons}. As demonstrated in Kitaev’s seminal paper, the TC phase hosts three distinct types of topological excitations—namely, the electric charge ($e$), magnetic charge ($m$), and a composite dyon ($\varepsilon=e\times m$). The nature and dynamics of these excitations provide a natural pathway toward topological quantum computation and fault-tolerant error correction. Interestingly, introducing perturbative magnetic fields induces non-trivial dynamics among these charges \cite{PhysRevB.82.085114, PhysRevLett.106.107203, PhysRevB.102.235124, PhysRevB.104.195115}. In this context, it is natural to ask whether the anisotropic limit of the quasicrystal also realizes a TC-like model, and if so, what are the properties of its ground state and low-lying excitations?

\begin{figure}
    \centering
 \includegraphics[width=1\columnwidth]{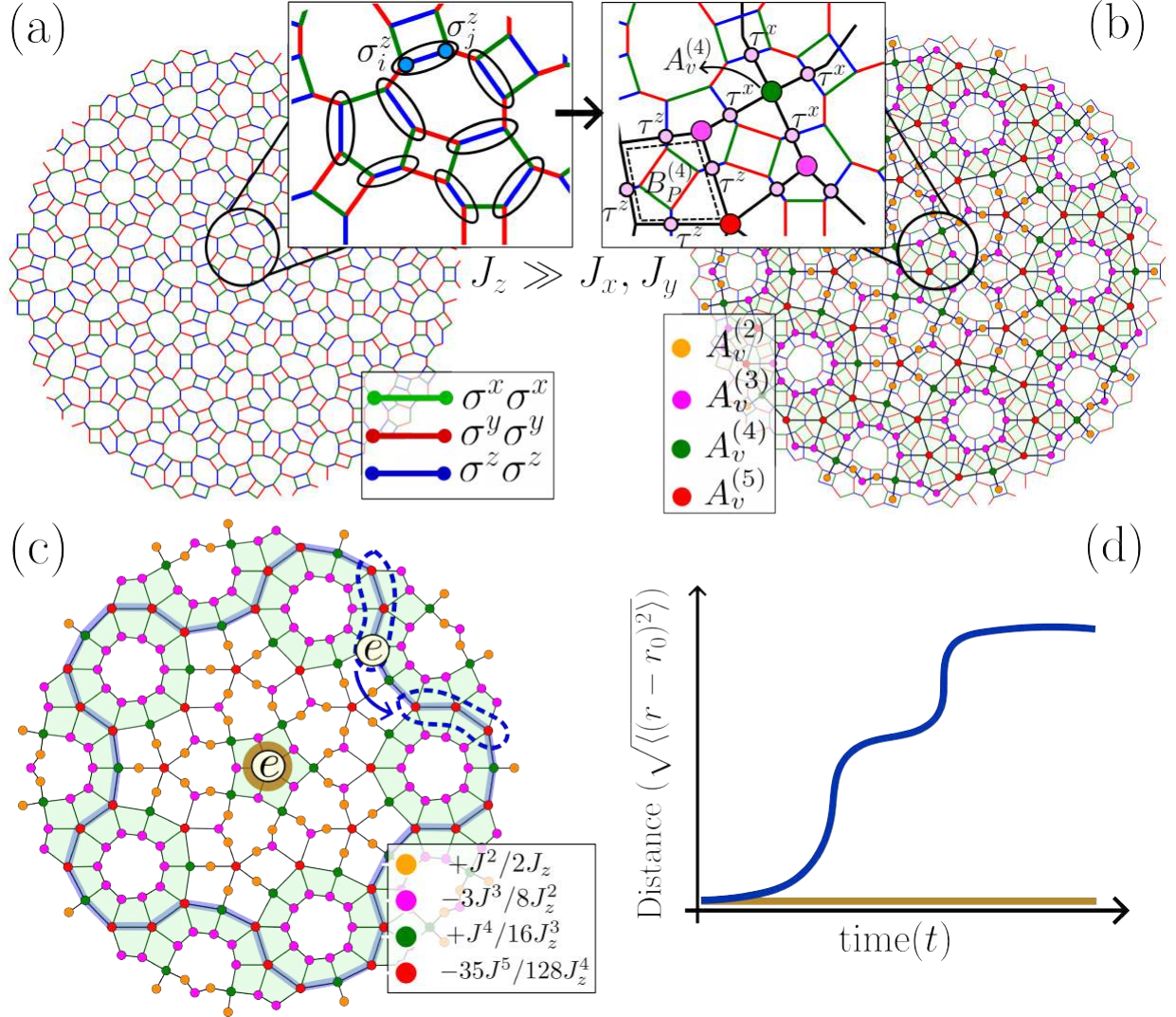}
\caption{{\bf Quasicrystalline Toric code:~} (a) Kitaev QSL on a tri-coordinated quasicrystalline lattice (shown for Gen 5). In the anisotropic limit ($J_{z}\gg J_{x}, J_{y}$), pairs of site-spins $\sigma^{\alpha}_{i,j}$ (blue circles, inset of (a)) are locked into bond-spins $\tau^{\alpha}$ (purple circles, inset of (b)). The corresponding low-energy Hamiltonian describing these bond-spins is given by the TC model (Eq.\eqref{Hamiltonian-TC}) (b) Vertices ($v$) and links defining the star operators ($A_{v}^{(n)}$) generate the effective TC lattice, with colors indicating the different coupling strengths ($\lambda_{A}^{(n)}$) of $A_{v}$. The plaquettes correspond to flux terms $B_{p}^{(n)}$ in Eq.\eqref{Hamiltonian-TC}, and the sign of $B_{p}$ couplings induces a ground state threaded by $\pi$ fluxes ($m$ anyons), indicated by green-shaded plaquettes. (c) Motion of an $e$ anyon in this quasiperiodic potential and $\pi$-flux landscape (Eq.\eqref{H-eff-boson}) shows either full localization, as shown for the central $e$ anyon, or quasi-localization, where the particle remains confined within small loops for extended times before tunneling into neighboring loops of similar structure. (d) The wavepacket spreading distance (Eq.\eqref{eqn-msd}) as a function of time displays either a flat profile (for localized cases) or step-like plateau behavior (for quasi-localized cases).} 
\label{Figure1}
\end{figure}

Our study begins with the tri-coordinated quasicrystalline (QC) spin liquid reported in Ref.~\cite{PhysRevB.110.214438}. By taking the strong-$z$ anisotropy limit, we demonstrate that, independent of the lattice generation (Gen), the resulting TC model exhibits a hierarchical structure of \textit{star} and \textit{plaquette} operators of various orders, accompanied by a hierarchy of exponentially separated energy scales. Notably, when this same methodology is applied to the honeycomb lattice, it yields a TC model that is translationally symmetric both in terms of operator content and coupling constants. We show that the effective QC toric code (QCTC) Hamiltonian possesses an intricate structure governed by the hierarchy of energy scales, which are intimately connected to the underlying QC geometry. Furthermore, the action of a perturbative magnetic field gives rise to qualitatively new phenomena absent in crystalline counterparts. In particular, states at the same energy can show both localized and extended characteristics, with their dispersion under increasing field strength $h$ displaying distinct scaling behaviors depending on the nature of the corresponding wavefunctions. Interestingly, anyonic charges in the QCTC are found to be either fully localized or undergo a stepwise delocalization process, where the charge spreads sequentially over subsets of sites of increasing size. We define this behavior as \textit{quasilocalization} \textcolor{blue}{(see Fig.\ref{Figure1}(c))}, akin to the phenomenon identified by Passaro \textit{et al.} in their seminal work on electron delocalization in quasicrystals under strong coupling conditions~\cite{PhysRevB.46.13751}. In addition, we find that beyond the energy hierarchy, the flux configurations in the ground state themselves induce interference effects that can localize anyonic states. These localized modes are immune to any perturbative magnetic field, arising purely from the interplay between the quasicrystal geometry and the intrinsic flux distribution. \textcolor{blue}{Finally, we classify the different types of isolated phason defects that can arise in our quasicrystalline lattice and examine how these defects locally alter the arrangement of TC stabilizer operators, thereby giving rise to nontrivial modifications of the ground-state.} 

In Sec.~\ref{sec:model}, we introduce the model and derive the effective Hamiltonian in the strongly anisotropic limit. Section~\ref{sec:groundst} discusses the ground-state properties. In Sec.~\ref{sec:zeeman}, we incorporate a magnetic field into the system and qualitatively describe the anyon dynamics and the resulting phases at different field strengths. Section~\ref{sec:sps} presents an analysis of the single-particle spectrum, while Sec.~\ref{sec:dyne} focuses on the dynamics of electric charges and their anomalous localization properties. In Sec.~\ref{sec:piflux}, we examine how regions of $\pi$ flux, together with geometric constraints, give rise to localized states. \textcolor{blue}{In Sec.~\ref{sec:phason}, we discuss the effects of elementary phason defects on the TC ground state.} Finally, Sec.~\ref{sec:conc} concludes the paper and outlines the future directions.

\section{Model}
\label{sec:model}

Our starting point is the Kitaev model \cite{kitaev2006anyons} defined on a tri-coordinated quasicrystalline lattice. This lattice is generated by connecting the centroids of the two fundamental isosceles triangles—the golden triangle and the golden gnomon—that form the building blocks of the five-fold rotation symmetric Penrose quasicrystal (see Supplementary material (SM) \cite{supplement-1} and Ref.\cite{PhysRevB.110.214438} for details). The spin-$\frac{1}{2}$ Pauli operators on each site of this tri-coordinated lattice couple each other via nearest-neighbor Ising-like interactions ($\sigma^{\alpha}_{i}\sigma^{\alpha}_{j}$), where the easy-axis ($\alpha$) depends on the bond-type: for green (G), red (R), and blue (B) bonds, $\alpha=x,y,z$ respectively (see Fig.\ref{Figure1}(a)). The Hamiltonian is given by,
\begin{align}
H_{K} = -\sum_{\alpha=x,y,z}J_{\alpha}\sum_{\langle ij\rangle_{\alpha}}\sigma^{\alpha}_{i}\sigma^{\alpha}_{j} \label{H-kitaev}
\end{align}
where the spin exchange couplings ($J_{\alpha}$) are chosen to be ferromagnetic (i.e. $J_{\alpha}>0$). Ref.\cite{PhysRevB.110.214438} showed that the ground state of this model is a gapped $\mathbb{Z}_{2}$ quantum spin liquid throughout the entire range of $J_{\alpha}$. Unlike the honeycomb Kitaev model, whose ground state is devoid of $\pi$ fluxes, the quasicrystalline variant energetically favors the presence of $\pi$ fluxes in certain plaquettes (squares and octagons). 

Here, we focus on the TC limit \cite{kitaev2006anyons} of the above Hamiltonian, which arises when one of the $J_{\alpha}$ couplings is much larger than the other two. Specifically, we choose $J_{z} \gg J_{x}, J_{y}$ \cite{coupling-choice-note}. The effective low-energy Hamiltonian in the large $J_{z}$ limit can be obtained from the standard degenerate perturbation theory in $J_{\alpha}/J_{z}$ ($\alpha=x,y$). Deferring the detailed derivation to the SM \cite{Supp-note}, we summarize below the essential steps involved in obtaining the effective TC Hamiltonian. We start from the extreme limit, when $J_{x}=J_{y}=0$. The zeroth-order Hamiltonian ($H_{0}$) describes a system of decoupled $zz$ bonds (see Fig.\ref{Figure1}(a) inset). Each of these $zz$ dimers has two-fold degenerate ground states (GSs): $|\uparrow \uparrow \rangle $ and $|\downarrow \downarrow \rangle $. This gives rise to an effective pseudo-spin-$\frac{1}{2}$ degree of freedom that can be thought of as residing on the centers of the $z$-bonds,
\begin{align}
\ket{\tau^{z}=+1} = \ket{\uparrow \uparrow}_{z}\ \ \ \text{and }\ \ \ket{\tau^{z}=-1} = \ket{\downarrow \downarrow}_{z}
\end{align}
For each strong $z$ bond (labeled by $l$), we introduce a pair of canonically conjugate Pauli pseudo-spin operators defined by
\begin{align}
\tau^z_{l}=\ket{\uparrow \uparrow}\bra{\uparrow \uparrow}-\ket{\downarrow \downarrow}\bra{\downarrow \downarrow}\nonumber\\
\tau^x_{l}=\ket{\uparrow\uparrow}\bra{\downarrow\downarrow}+\ket{\downarrow\downarrow}\bra{\uparrow\uparrow}
\end{align}
which, respectively, encode the bond-spin configurations and the transitions between the two possible bond-spin states. When $J_{x}$ and $J_{y}$ are switched on perturbatively, the massive GS degeneracy of $H_{0}$ is lifted by transitions between states within the GS subspace at various orders (of $J_{\alpha}/J_{z}$) and results in the effective TC Hamiltonian which has multi-spin interactions of the bond-spins ($\tau^{\alpha}$). Following the standard convention \cite{kitaev2006anyons}, we assign the $\tau^{\alpha}$ spins to the links of the \textit{dual} lattice, such that each dual link intersects a strong $z$ bond of the original tri-coordinated lattice (see Fig.\ref{Figure1}(b), also see SM). On this dual TC lattice, the Hamiltonian comprises two kinds of operators: (1) \textit{star} ($A_{v}^{(n)}$) operators, located at the vertices ($v$) of coordination number $n$ of the TC lattice, 
\begin{align}
 A^{(n)}_{v}=\prod_{l(v)=1}^{n}\tau^{x}_{l}   
\end{align}
where $l(v)$ corresponds to links that emanates from the vertex $v$ of the dual TC lattice. (2) The \textit{plaquette} ($B_{p}^{(n)}$) operators, defined on the minimal plaquettes $p$ (having $n$ sides) of the dual lattice, are given by
\begin{align}
  B^{(n)}_{p}=\prod_{l(p)=1}^{n}\tau^{z}_{l}  
\end{align}
where $l(p)$ denotes the links of the plaquette $p$. These operators are Ising-like: flipping their eigenvalues from $+1$ to $-1$ corresponds to the creation of an $e$ anyon (for $A_{v}^{(n)}$) or an $m$ anyon (for $B_{p}^{(n)}$) at the corresponding vertex $v$ or plaquette (center) $p$, respectively.  The Hamiltonian is given by
\begin{align}
H_{\text{TC}}=\sum_{v}\lambda_A^{(n)}A_{v}^{(n)}+\sum_{p}\lambda_{B}^{(n)}B_{p}^{(n)} \label{Hamiltonian-TC}
\end{align}
The coupling parameters $\lambda^{(n)}_{A/B}$ of $A_{v}$ and $B_{p}$ operators are obtained from the perturbation theory calculations (see SM \cite{Supp-note}). For the $A_{v}$ terms, only the couplings corresponding to $n = 2, 3, 4, 5$ are nonzero, and to the leading order \cite{perturbation-note} these are given by, $\lambda^{(2)}_{A} = +J^{2}/2J_{z}$, $\lambda^{(3)}_{A}= -3J^{3}/8J_{z}^{2}$, $\lambda^{(4)}_{A}= +5J^{4}/16J_{z}^{3} $, and $\lambda^{(5)}_{A}= -35J^{5}/2^{7}J_{z}^{4} $. For $B_{p}$ operators, the non-vanishing coupling parameters are: $\lambda^{(4)}_{B}=+J^{4}/16J_{z}^{3}$, $\lambda_{B}^{(6)}=-3J^{6}/2^{8} J^{5}_{z}$, and $\lambda_{B}^{(10)}=-3J^{10}/2^{16}J^{9}_{z}$. Here, we have chosen $J_{x}=J_{y}=J$, see SM for the general expressions. All the operators present in the Hamiltonian commute with each other, which makes the model exactly solvable in absence of any perturbations. The model also preserves the time-reversal (TR) invariance of $H_K$ (Eq.\eqref{H-kitaev}) \cite{note-time-rev}.  

\section{Zero-field Ground State}
\label{sec:groundst}

In contrast to the square-lattice TC model, which contains only a single type ($n=4$) of $A_{v}$ and $B_{p}$ operators with a uniform sign structure, the QC version admits multiple values of $n$ and both positive and negative coupling parameters. This has important consequences for both the zero and finite Zeeman field ground states. The diversity in operators and coupling constants arises from the presence of plaquettes of various sizes in the underlying Kitaev quasicrystal, whose densities in the thermodynamic limit are irrational numbers, and scale as different powers of the golden ratio $\varphi = (\sqrt{5}+1)/2$ \cite{PhysRevB.110.214438},
\begin{align}
(\rho^{K}_{4},\ \rho^{K}_{6},\ \rho^{K}_{8},\ \rho^{K}_{10}) = \bigg(\frac{1}{\varphi^2}, \frac{1}{\varphi^2}, \frac{1}{\varphi^5}, \frac{1}{\varphi^4}\bigg)
\end{align}
where $\rho_{n}^{K}$ denotes the density of plaquettes with $n$ sides of the Kitaev quasicrystal. Since each of the Kitaev lattice plaquettes contributes to either $A_{v}$ or $B_{p}$ terms, the density scalings of the latter terms are given by (see SM \cite{supplement-2})
\begin{align}
&(\rho^{A}_{2},\ \rho^{A}_{3},\ \rho^{A}_{4},\ \rho^{A}_{5}) \approx \bigg(\frac{1}{2\varphi^{2}} ,  \frac{2}{3\varphi^{2}},\ \frac{1}{\varphi^{5}},\ \frac{2}{3\varphi^{4}}\bigg)\\
&(\rho^{B}_{4},\ \rho^{B}_{6},\ \rho^{B}_{10}) \approx \bigg(\frac{1}{2\varphi^{2}}, \frac{1}{3\varphi^{2}}, \frac{1}{3\varphi^{4}}\bigg)
\end{align}
Here, $\rho^{A(B)}_{n}$ denotes the density of $A_{v}^{(n)}$ (and $B_{p}^{(n)}$). The zero-field ground state of Eq.\eqref{Hamiltonian-TC} contains a finite density of $e$ anyons: these reside on the vertices of coordination numbers $n=2$ and $4$, for which $\lambda_{A}^{(n)}>0$. Likewise, $m$ anyons occupy the square ($n=4$) plaquettes of the TC lattice, since $\lambda_{B}^{(4)}>0$. Thus, the zero-field GS hosts a finite (irrational number) density of both $e$ and $m$ charges purely due to geometric reasons, which are given by,
\begin{align}
\rho_{e} \approx \frac{1}{2\varphi^{2}}+\frac{1}{\varphi^{5}}\ \ ,\ \ \rho_{m}\approx \frac{1}{2\varphi^{2}}
\end{align}
Interestingly, the vertices and plaquettes (of the TC lattice) that host $e$ and $m$ anyons in the zero-field ground state originate from the square and octagonal plaquettes of the Kitaev lattice. The latter ones also host $\pi$ fluxes in the isotropic limit ($J_{x}=J_{y}=J_{z}$) \cite{PhysRevB.110.214438}. Since the spin liquid is gapped throughout the full phase space of parameters, the finite-anyon-density ground state of the TC phase is adiabatically connected to the isotropic-limit ground state with static $\pi$ fluxes.

The model has only $\mathbb{Z}_{2}$ (parity) symmetry for the anyons; consequently, when Zeeman fields (or other perturbations) are introduced, the low-energy dynamics of one anyon species gets affected solely by the finite density of the other species, and not by its own kind. The effects of background $\rho_{m}$ on the low-energy dynamics of an $e$ particle (or hole) are discussed in Sec.\ref{sec:piflux}.  The topological properties (such as the GS degeneracy) remain unaffected by the quasicrystalline geometry, provided that periodic boundary conditions are properly implemented \cite{ft-note-pbc}.

\begin{figure}
    \centering
\includegraphics[width=1.0\columnwidth]{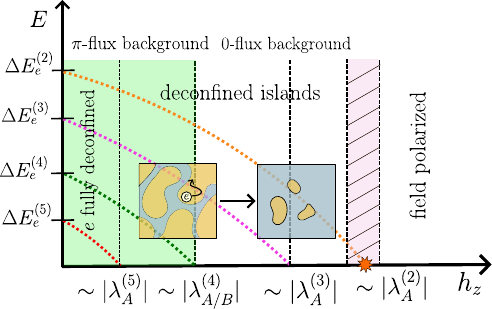}
\caption{{\bf Schematic phase diagram of $e$ anyon condensation transition:~} The generation of $A_{v}^{(n)}$ at different orders ($n$) of the perturbation theory results in an exponentially separated energy gap scales ($\Delta E_{e}^{(n)}$) for the $e$ anyon excitations. With increasing Zeeman field, $\Delta E_{e}^{(n)}$ decreases and expected to become zero at a field strength $\sim O(\lambda_{A}^{(n)})$. These energy scales demarcate the boundaries of the different phases as discussed in the main text.}
\label{Figure2}
\end{figure}

\section{Action of Zeeman Fields}
\label{sec:zeeman}

We now consider the effect of perturbative Zeeman field on $H_{TC}$ (Eq.\eqref{Hamiltonian-TC}),
\begin{align}
H_{Z} = -h_{z}\sum_{l}\tau^{z}_{l} - h_{x}\sum_{l}\tau^{x}_{l}
\end{align}
In terms of the original spin degrees of freedom, $\tau^{z}_{l}\sim (\sigma^{z}_{i}+\sigma^{z}_{j})/2$ and $\tau^{x}_{l}\sim \sigma^{x}_{i}\sigma^{x}_{j}$, where $l$ denotes the dual link corresponding to the link $(ij)$ of the tri-coordinated lattice. Consequently, $\tau^{z}$ breaks TR symmetry and corresponds to the magnetic-field coupling in the $z$-direction of the original Kitaev model. In the following discussion, we consider the effects of this coupling. The $\tau^{x}$ coupling, on the other hand, preserves TR and such terms are generated in the second-order ($O(\tilde{h}_{x}^{2}/J_{z})$) perturbation theory when the Kitaev model is acted by $-\tilde{h}_{x}\sum_{i}\sigma^{x}_{i}$ \cite{PhysRevB.102.235124, PhysRevB.104.195115}. When $h_{x,z}=0$, the excitations are gapped. They can be labeled as particle- or hole-like excitations, depending on whether the couplings ($\lambda_{A,B}$) are negative or positive. Since $\tau^{z}$ anti-commutes with $A_{v}$, local $e$-parity conservation is lost, and $e$ anyons become dynamic. On the other hand, $[\tau^{z}_{l}, B_{p}^{(n)}]=0$. Therefore, $m$ anyons do not acquire any dynamics. 

The zero-field energy gap for the $e$ anyons is position-dependent and determined by the coordination number ($n$) of the site on the TC lattice where the excitation is created [Fig.\ref{Figure1}(c)]. Depending on $n$, there exist four exponentially separated gap scales, $\Delta E_{e}^{n} = 2|\lambda_{A}^{(n)}|$, corresponding to $n = 2, 3, 4, 5$. When $h_{z}$ becomes nonzero, the mobile $e$ anyons experience a non-uniform potential landscape set by these gap scales. Their dynamics is governed by two processes: free hopping within spatially disconnected equipotential regions (or contours), and weak tunneling across potential barriers (or wells) arising from the spatial variation in $\Delta E_{e}^{(n)}$.

Let us qualitatively discuss the different phases that arise as the field strength ($h_{z}$) is varied. As $h_{z}$ increases, the excitation gaps gradually shrink and eventually close when $h_{z}\sim O(|\lambda_{A}^{(n)}|)$ [see Fig.\ref{Figure2}]. Near the first gap-closing point (around $|\lambda_{A}^{(5)}|$), the density of $e$ anyons increases within the corresponding degenerate patches, eventually leading to Bose condensation—equivalently, $\langle \tau^{z}_{l} \rangle \neq 0$ for links ($l$) within these regions \cite{Senthil-Fisher, PhysRevLett-trebst-toric-zeman, wu-prokofev-prb, shifeng-adhip-prb}. The $e$ anyons residing in the higher-energy ``plateaus'' still remain gapped, 
resulting in the system fragmenting into disconnected puddles where the excitations remain deconfined, separated by ``lakes'' in which $e$ particles have Bose-condensed. With further increase in $h_{z}$, these puddles shrink, and when $h_{z} \sim O(|\lambda_{A}^{(2)}|)$, the system transitions into a trivially field-polarized phase devoid of stable anyonic excitations. The static background of $m$ anyons ($\pi$-fluxes) present at zero and small Zeeman fields is also expected to vanish once $h_{z} \gtrsim O(|\lambda_{B}^{(4)}|)$, as a large $h_{z}$ favors a $\pi$-flux-free configuration.

\begin{figure*}
    \centering
    \includegraphics[width=1.0\textwidth]{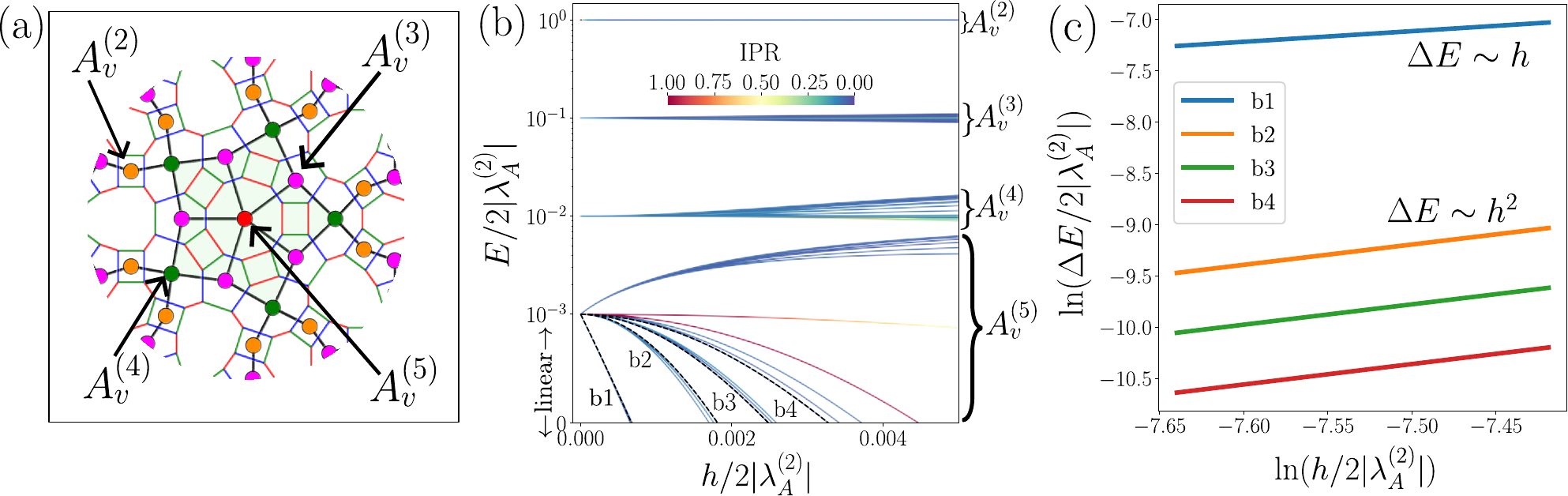}
    \caption{{\bf Low-energy eigenstates of the $e$ anyon hopping model (Eq.\eqref{H-eff-boson}):~} (a) The star operators ($A_v^{(n)}$) of different orders shown in the zoomed version of the TC lattice. (b) Energy spectrum ($E$) of the Gen-5 lattice as a function of Zeeman coupling, $h$ (for $g=J/J_{z}=0.1$) consists of four exponentially separated bands associated with $e$ excitations that are initially localized (when $h=0$) at the vertices of different coordination numbers (corresponding to different $A_{v}^{(n)}$ in Fig.(a)). The color map depicts the IPR (Eq.\eqref{ipr-eqn}) for a given $E$ and $h$. The $E$ axis is linear up to $E\sim 10^{-3}$ (in units of $2|\lambda_{A}^{(2)}|$), after that, it is shown in the log-scale for clear visualization of the different bands. (c) Scaling of the $e$ anyon gap ($\Delta E$) with $h$, for a few low-lying states of the lowest band, shows mostly quadratic scaling ($\sim h^{2}$), except for the lowest branch (b1), which scales $\sim h$. }
    \label{Figure3-part1}
\end{figure*}

In the square-lattice TC model, the $e$-anyon condensation transition is known to be second order, with the universality class determined by the background flux configuration. Specifically, it belongs to the 3D Ising universality class in the zero-flux case \cite{vidal-dussel-schmidt, PhysRevB.85.195104, PhysRevB.100.125159}, and to the 3D XY universality class in the presence of a dense $\pi$-flux background \cite{PhysRevB.44.686, PhysRevB.65.024504, PhysRevLett.109.187202}. These results follow from the exact mapping of the TC model onto the two-dimensional transverse-field Ising model (TFIM) with zero and $\pi$-flux backgrounds, respectively. In our QCTC model, a similar mapping to an effective TFIM can still be constructed (see below). However, due to the lack of translational symmetry and limited accessible system sizes, it remains challenging to determine the precise nature and universality classes of the phase transitions separating different phases. 

Mapping to TFIM \cite{PhysRevLett.98.070602, PhysRevB.100.125159} goes as follows: Since the star ($A_{v}$) operators are Ising-like, these can be represented as a dual Pauli spin operator at each of the vertices, $A_{v}^{(n)}\mapsto \eta^{z}_{v}$. The Zeeman field ($\tau^{z}_{l}$) either moves an $e$ anyon to its neighboring vertices or creates a pair of $e$ anyons out of the vacuum. Therefore, $\tau^{z}_{l}\mapsto \eta^{x}_{v}\tau^{z}_{vv'}\eta^{x}_{v'}$, where $l$ corresponds to the link that connects two nearest-neighbor vertices $v$ and $v'$. The link variables $\tau^{z}_{vv'}$ keep track of the background $\pi$-flux (static $m$ anyon) distribution within the plaquettes $p_{n}$ with $n$ edges of the TC lattice. For field strength $h_{z}< O(|\lambda_{B}^{(4)}|)$, the 4-sided plaquettes of the TC lattice host $\pi$-fluxes, in the low-energy subspace (see Fig.\ref{Figure1}). Therefore, 
\begin{align}
\prod_{\langle vv'\rangle\in p_{n}}\tau^{z}_{vv'}\ket{\psi_{\text{phy}}}= -\text{sgn}(\lambda_{B}^{(n)})\ket{\psi_{\text{phy}}}\label{flux-constraint}
\end{align}
where $\ket{\psi_{\text{phy}}}$ corresponds to the low-energy physical states of the model when $h_{z}< O(|\lambda_{B}^{(4)}|)$. For $h_{z} > |\lambda_{B}^{(4)}|$, the flux background is no longer favored at low energies, and $\tau^{z}_{l}\sim \eta^{x}_{v}\eta^{x}_{v'}$, which corresponds to ordinary anyon hopping and pair creation without any phase accumulation.

Here, we restrict ourselves to the dynamics of $e$ anyons for perturbative field strengths ($h_{z}\ll |\lambda_{A}^{(5)}|$), in which case the anyons are well-defined gapped excitations throughout the lattice with background $\pi$ fluxes. The effective low-energy model is given by
\begin{align}
H_{\text{TFIM}}=\sum_{v}\lambda_{v}^{(n)} \eta^{z}_{v} -h_{z}\sum_{\langle vv'\rangle}\eta^{x}_{v}\tau^{z}_{vv'}\eta^{x}_{v'}
\end{align}
subjected to the flux constraints Eq.\eqref{flux-constraint}. The transverse field $\lambda_{v}^{(n)}$ depends on a vertex $v$ solely through its coordination number $n$, and $\lambda_{v}^{(n)}= \lambda_{A}^{(n)}$. In the following, our focus will be on understanding the low-field single-particle spectrum and localization dynamics of this effective model.

\section{Single-particle Spectrum}
\label{sec:sps}

\begin{figure*}
    \centering
    \includegraphics[width=1.0\textwidth]{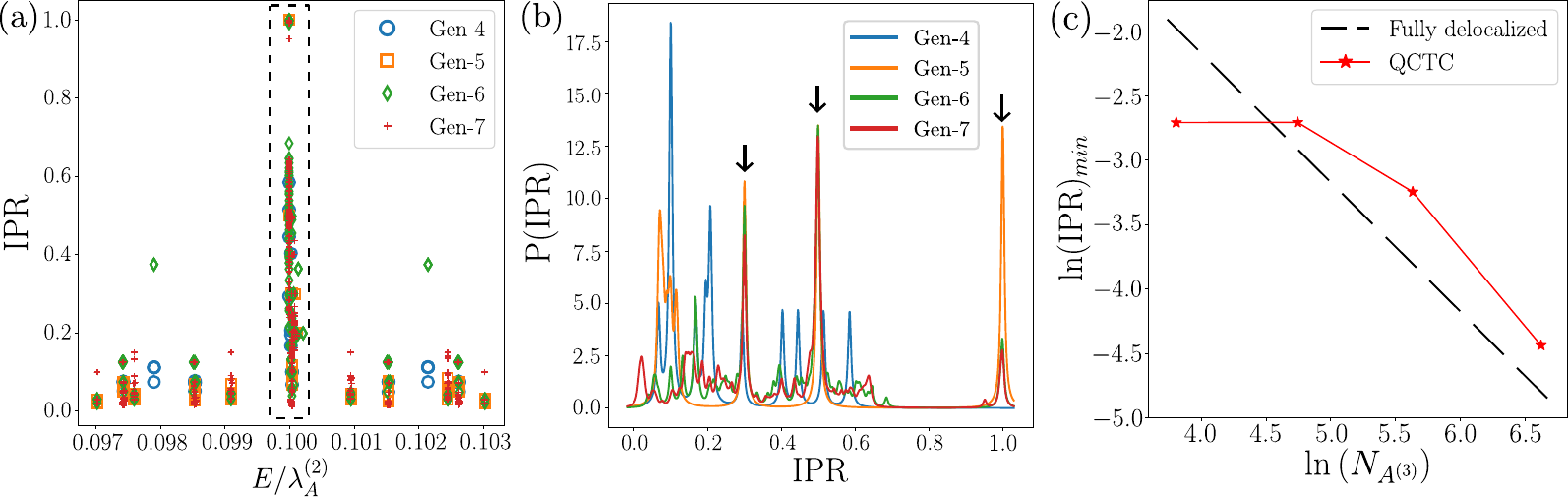}
    \caption{{\bf Localization properties of the Eigenstates:} \textcolor{blue}{(a) IPR vs. energy $E$, for the $A_{v}^{(3)}$ band (centered at $E/2|\lambda_{A}^{(2)}|=0.1$) shown for $h=0.0015$, and for four generations with increasing system size. For a given $E$, the states have extremely varying localization properties, as shown in Fig.(b), where a Lorentzian-smoothed probability distribution of IPR (P(IPR)) is plotted for $E\sim 0.1$. Special IPR values are marked by arrows (see main text). (c) Scaling of IPR, for lowest-IPR states as a function of $N_{A^{(3)}}$ (= number of $A^{(3)}$-type sites) is compared with the fully delocalized states, for which IPR $\sim 1/N$.}}
    \label{Figure3-part2}
\end{figure*}

We transform Pauli spins into hardcore bosons (\cite{sachdev_QPT}) by the mapping: 
$\eta_{v}^{z}=2b^{\dag}_{v}b_{v}-1,\ (\text{for }\lambda_{v}^{(n)}>0)$, $\eta_{v}^{z}=1-2b^{\dag}_{v}b_{v}\ (\text{for }\lambda_{v}^{(n)}<0)$. This ensures that the excitations have zero-field energy gap $2|\lambda_{v}^{(n)}|$, irrespective of whether they are particle- or hole-like. The $\tau^{x}_{v}$ creates or annihilates an excitation; therefore,
$\eta^{x}_{v}= b_{v}+b_{v}^{\dag}$. Implementing this mapping and restricting ourselves to the single $e$-anyon subspace, we obtain the following simplified boson hopping model, 
\begin{align}
H_{b}= \sum_{v}\lambda_{v}^{(n)}\text{sgn}(\lambda_{v}^{(n)})(2b_{v}^{\dag}b_{v}-1) - h\sum_{\langle vv'\rangle}(b^{\dag}_{v}b_{v'}+\text{H.c.}) \label{H-eff-boson}
\end{align}
Hereafter, we drop the subscript of the Zeeman field and denote it by $h$. From now on, we also ignore the precise numerical factors present in $\lambda_{v}^{(n)}$ (or $\lambda_{A}^{(n)}$), and take $|\lambda_{v}^{(n)}|=J_{z}g^{n}$, where $g = J/J_{z}$, and $n=2,3,4,5$. Since $H_{b}$ is quadratic, the properties of its low-energy eigenstates can be easily determined from exact diagonalization of various generations. The results are summarized in Fig.\ref{Figure3-part1} and Fig.\ref{Figure3-part2}, and described as follows.

The energy spectrum of the model on the generation-5 lattice (251 sites), plotted as a function of the perturbative field $h$, reveals four well-separated energy bands centered around $E \sim 2|\lambda_{A}^{(2)}| g^{n}$ and $n = 0, 1, 2, 3$ (Fig.\ref{Figure3-part1}(b))~\cite{spectrum-note}. These bands correspond to single $e$-anyon excitations residing on TC lattice sites with different coordination numbers, each associated with the flipping of $A_{v}^{(n)}$ operators of different orders (see Fig.~\ref{Figure3-part1}(a)). Due to the spatial non-uniformity in the onsite potentials (or couplings of $A_{v}^{(n)}$), the energy dispersion with $h$ exhibits both linear ($\sim h$) and quadratic ($\sim h^{2}$) scaling, as illustrated for selected low-lying states from the $A^{(5)}$ sector in Fig.~\ref{Figure3-part1}(c)~\cite{uniform-potential-comment}.

For the $A^{(5)}$ and $A^{(3)}$ sectors, the presence of equipotential nearest-neighbor (NN) sites coupled via hopping gives rise to the linear dependence on $h$. In contrast, when such equipotential NN connections are absent, the leading contribution arises from virtual hopping through high-energy sites, resulting in $O(h^{2})$ scaling. Higher-order ($h^{m}$ with $m>2$) scaling is generally suppressed unless quadratic corrections cancel, which requires fine-tuning \cite{ring-model-note}. The $O(h^{2})$ scaling can arise from fully localized states, where a particle briefly hops out of the localized region and returns, leading to a second-order process. In contrast, extended states exhibiting the same scaling originate from a non-trivial interplay between phase fluctuations and hybridization among sites.

To obtain a quantitative understanding of the localization properties of the eigenstates, we calculate the inverse participation ratio (IPR) in the position basis, which is defined as 
\begin{align}
\text{IPR} = \sum_{i}|\psi_{i}|^{4} \label{ipr-eqn}
\end{align}
where, $\psi_{i}$ is the probability amplitude at the $i$-th site of the TC lattice. The value of IPR is shown in Fig.\ref{Figure3-part1}(b), along with the $h$ dependence of the spectrum. For every $A^{(n)}$ sector, the eigenstates exhibit a mixture of strongly localized (IPR $\sim 1$) and extended-like (low IPR) behavior. This arises because the equipotential sites within each sector are not distributed uniformly across the lattice: some occur in close proximity, allowing the wavefunction to delocalize through nearest-neighbor hopping ($O(h)$ scaling), whereas others are separated by larger distances, forcing the wavefunction to rely on higher-order hopping processes and thereby producing strong localization features. Additionally, the combined effects of background $\pi$-flux pattern and geometric constraints sometimes enhance localization (see below).

To quantify these localization properties, we focus on the $A^{(3)}$ band (centered around $E/2|\lambda_{A}^{(2)}|=0.1$), which has the largest number of states for any given generation. As shown in Fig.~\ref{Figure3-part2}(a), this band contains both localized and extended states within a narrow energy window, independent of the generation (i.e., system size). Notably, around the band center ($E/2|\lambda_{A}^{(2)}|\approx 0.1$), the IPR values spread over a wide range. This is very different from the usual Anderson-like disorder systems, where the localization length ($\xi(E)$) is usually a single-valued function of single-particle excitation energy, $E$ \cite{PhysRevB.109.014210, PhysRevA.109.013314}.

In Fig.~\ref{Figure3-part2}(b), we analyze the Lorentzian-broadened probability distribution of IPR, denoted as $P(\mathrm{IPR})$, within a small energy window near $E/2|\lambda_{A}^{(2)}|=0.1$. Across all four generations studied, the distribution consistently shows three peaks (\textcolor{blue}{arrows in Fig.\ref{Figure3-part2}(b)}) located near IPR values of $1$, $1/2$, and $1/3$—indicating eigenstates concentrated on approximately one, two, and three sites, respectively. In contrast, the low-IPR portion of the distribution shows no universal features and varies with generation. Nevertheless, as the generation increases, the position of the lowest observed IPR peak shifts progressively to smaller IPR values, determining the scaling of extended states with system size. Fig.\ref{Figure3-part2}(c) plots $\mathrm{IPR}_{\mathrm{min}}$ for the $A^{(3)}$ sector against the number of $A^{(3)}$-type sites, $N_{A^{(3)}}$. As the generation number (system size) increases, the scaling trend approaches the $1/N$ behavior expected for fully delocalized states. A more extensive system-size analysis is required to make definitive statements about the thermodynamic limit behavior, which lies beyond the scope of this work.

\begin{figure}
\includegraphics[width=1\columnwidth]{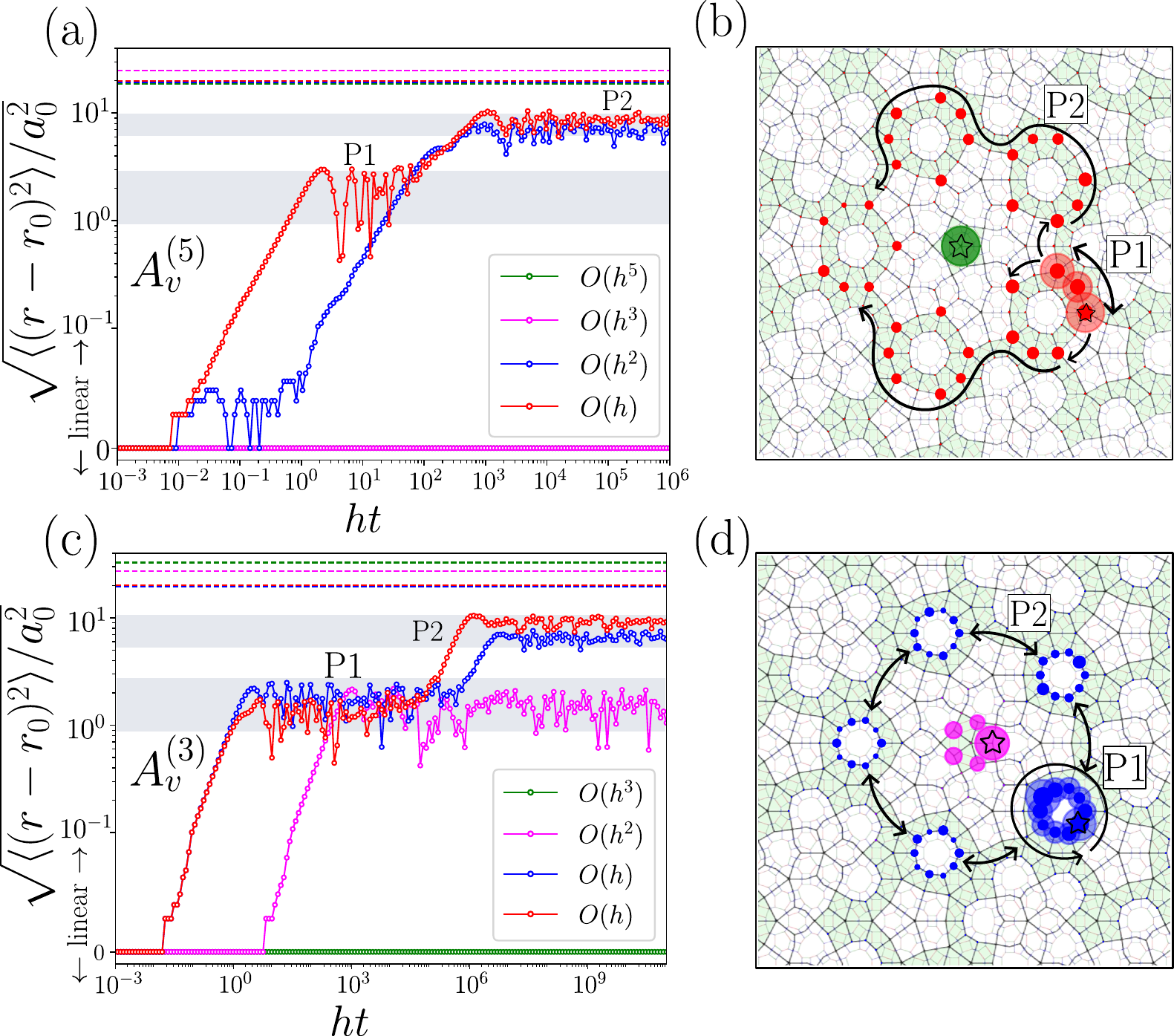}
\caption{{\bf Spreading of an $e$ anyon wavepacket over time:~} Mean-square displacement $d(r_{0},t)$ (Eq.\eqref{eqn-msd}) of an $e$ anyon wavepacket as a function of time ($t$) for two different initial conditions. The average lattice spacing is denoted by $a_{0}$, and the Zeeman coupling is set to $h = 10^{-4}\lambda_{A}^{(2)}$ (with $\lambda_{A}^{(2)} = 1$). Panels (a) and (c) correspond to initial wavepackets placed at $A^{(5)}$ and $A^{(3)}$ vertices, respectively; insets show the grouping of vertices by their hopping distances. In both cases, the wavepacket either remains fully localized, yielding a flat plateau, or undergoes discrete plateau transitions as it spreads across contours of increasing size, with exponentially long residence times on each plateau. Dashed lines indicate the mean displacement for wavepackets fully delocalized over equivalent vertex types. Panels (b) and (d) illustrate the time-averaged probability densities at two plateaus (P1 and P2) for quasi-localized cases (red, blue) and over the entire evolution for fully localized cases (green, purple). Initial wavepacket positions and their evolution are marked by star symbols and probability densities with different colors. Black arrows illustrate the movement of the anyon wavepacket. All data shown here is for Gen-7 lattice.}
\label{Figure4}
\end{figure}

\section{Dynamics of $e$ Anyons}
\label{sec:dyne}

Given the rich and non-trivial localization characteristics of the single-$e$ anyon eigenstates, we next examine the real-time dynamics of these quasiparticles. Specifically, we study the time evolution of an $e$-anyon wavepacket (equivalently, a bosonic wavepacket of Eq.~\eqref{H-eff-boson}) that is initially localized at a single lattice site $r=r_{0}$ of the TC lattice. To quantify the spatial spread of the wavepacket, we compute the time-dependent mean-square displacement, $d(r_{0},t)$~\cite{PhysRevB.62.15569}, defined as
\begin{align}
d(r_{0},t)\equiv \sqrt{\langle (r-r_{0})^{2}\rangle}
= \left[ \sum_{i} (r_{i}-r_{0})^{2}|\psi_{i}(t)|^{2} \right]^{1/2} \label{eqn-msd} 
\end{align}
where $\psi_{i}(t)=\langle i| e^{-iH_{b} t}|\psi^{(b)}(t=0)\rangle$ denotes the wavefunction amplitude at site $i$ at time $t$. Since the displacement implicitly depends on the location of the initial wavefunction, we need a classification scheme of the vertices of the TC lattice. Within each $A^{(n)}$ sector, we group the vertices according to a quantity we call the \textit{hopping distance}, which measures the minimum perturbative order in the Zeeman coupling $h$ needed to connect any two equipotential sites. Intuitively, wavepackets initialized at sites with a larger hopping distance—i.e., sites that are only connected through higher-order processes $O(h^{m})$ with large $m$—are expected to spread much more slowly across the lattice. We specifically focus on wavepackets belonging to the $A^{(5)}$ and $A^{(3)}$ sectors, because in the TC lattice, only these sectors have $O(h)$ sites, \textcolor{blue}{across all generations}.   

\begin{figure*}
    \centering
    \includegraphics[width=1.0\textwidth]{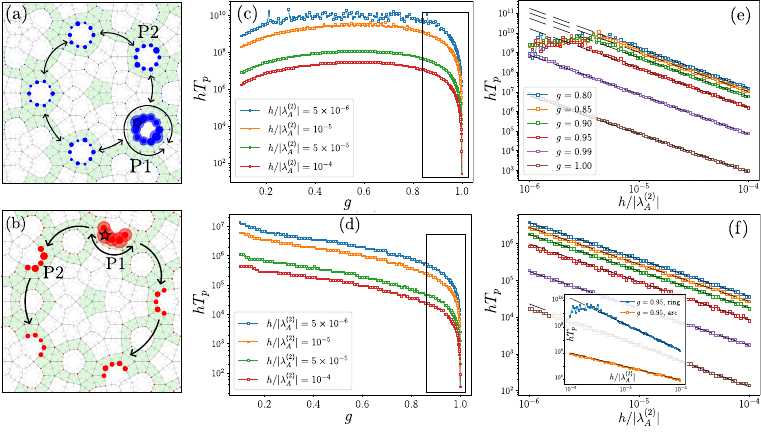}
    \caption{\textcolor{blue}{Dependence of plateau residence time ($T_p$) on Zeeman field strength ($h$) and the anisotropy in $A_v$ couplings, $g=|\lambda_A^{(n+1)}/\lambda_A^{(n)}|$: Wavepacket dynamics for two different $A^{(3)}_v$-type ($O(h)$) sites of Gen-7 TC quasicrystal are shown in (a) and (b). The initial site-localized wavefunction (shown by star symbols) spreads at first over the equipotential contours (P1) and stays there for the time interval $T_p$ before spreading across equivalent contours (P2). For these two wavepacket evolutions, the dependence of $T_p$ on $g$ for different values of $h/|\lambda_A^{(2)}|$ is shown in (c) and (d). In the strong-coupling regime (small $g$), the behavior is initial-state dependent, and in the weak-anisotropy limit ($g\rightarrow 1$), $T_p$ depicts a monotonic power-law decrease $T_p\sim (1-g)^{\alpha_p}$ (inside the boxed areas). The power-law dependence of $T_p$ on perturbative $h/|\lambda_A^{(2)}|$ for these two wavepacket dynamics, for different $g$ values (close to $g=1$), is shown in (e) and (f). In the inset of Fig.(f), we compare the two different $h$-dependent scalings of $T_p$ for the ring-type (Fig.(a)) and arc-like (Fig.(b)) wavepacket dynamics at a fixed $g=0.95$.}}
    \label{fig:plateau-res-time-scaling}
\end{figure*}

We now turn to the results shown in Fig.\ref{Figure4}. Throughout our analysis, we fix the Zeeman coupling to a small value, $h=10^{-4}|\lambda_{A}^{(2)}|$, ensuring that the excitations remain gapped. As illustrated in Fig.\ref{Figure4}(a) and (c), for both the $A^{(5)}$ and $A^{(3)}$ sectors, wavepackets initialized at $O(h)$ sites exhibit a sequential delocalization: the wavepacket spreads over a series of spatial contours (or patches), each of progressively larger size. Crucially, the wavepacket remains confined to each contour for an exponentially long duration. This results in the characteristic plateau-like features in the mean-square displacement $d(r_{0},t)$. \textcolor{blue}{Besides the dependence on the position of the initial state, the duration of each plateau ($T_p$) is controlled by two key parameters}:
(1) the anisotropy of the onsite energies, $g=|\lambda_{A}^{(n+1)}|/|\lambda_{A}^{(n)}|\approx J/J_{z}$, and (2) the \textcolor{blue}{effective} hopping \textcolor{blue}{strength $h/|\lambda_A^{(2)}|$} (see below). The wavepacket first spreads over the \textcolor{blue}{nearby} equipotential regions connected by $O(h)$ hopping, which typically spans only a few lattice spacings, on a time scale set by $h$. Subsequently, it tunnels between equivalent contours through potential barriers, a process that occurs on exponentially longer time scales. \textcolor{blue}{Similar sequential delocalization has also been observed in the wavepacket dynamics of other quasi-periodic Hamiltonians \cite{Theim-prb-rg, Theim-prb-qp-rg-chain}}. On the other hand, the sites connected by $O(h^{n})$ hopping with $n>1$ mostly remain localized within a few lattice spacings, but depending on the local environment, it can sometimes spread over larger distances. For example, the $O(h^{2})$ sites in the $A^{(5)}$ group are more dispersive than the $A^{(3)}$ group (see Fig.\ref{Figure4}).

In Fig.~\ref{Figure4}(b) and (d), we evaluate the probability density averaged over individual plateau residence times—equivalently, the temporal auto-correlation function~\cite{PhysRevB.62.15569},
\begin{align}
C(r_{i},r_{0}) = \frac{1}{T_{P_{n}}}\int_{t_{i}^{(P_{n})}}^{t_{f}^{(P_{n})}} |\psi_{i}(r_{0},t)|^{2} dt
\end{align}
to explicitly demonstrate the sequential delocalization dynamics. Here, $t_{i}^{(P_{n})}$ and $t_{f}^{(P_{n})}$ denote the beginning and end of the $n$-th plateau in the mean-square displacement evolution. As shown in Fig.~\ref{Figure4}(b), an $e$-anyon initialized on an $O(h)$ $A^{(5)}$-type site (indicated by a star symbol) remains confined to a local three-site chain during the P1 plateau. Only after a long tunneling time does the wavefunction leak out through higher-energy intermediate sites, gradually spreading over a larger contour \textcolor{blue}{(see also \cite{extra-wavepacket-plots})}. A similar sequential delocalization process occurs in the $A^{(3)}$ sector (Fig.~\ref{Figure4}(d)): during the first plateau, the wavepacket remains trapped within a small circular contour for a duration $t_{P_{1}}$, and only at later times ($t_{P_{2}}$) does it expand to a larger contour. Remarkably, such constrained \textit{quantum wheel}-like dynamics of low-energy quasiparticles has also been observed in interacting bosonic models on quasicrystalline lattices~\cite{PhysRevLett.131.173402}. This restricted spreading can be viewed as reminiscent of a Bose glass phase~\cite{PhysRevLett.126.110401}, originally identified in disordered Bose–Hubbard systems as an intermediate phase between Mott insulator and superfluid regimes~\cite{PhysRevB.40.546, PhysRevLett.103.140402}. However, because our model lacks $U(1)$ symmetry, the notion of compressibility used to diagnose the Bose glass does not apply, and thus the analogy is only qualitative.

\textcolor{blue}{In general, the dependence of $T_p$ on $g$ and $h/|\lambda_A^{(2)}|$ is nonuniversal in the strong-coupling regime (small $g$), where the functional form may differ even among initial states belonging to the same class. In Fig.\ref{fig:plateau-res-time-scaling}, we illustrate this feature using two distinct $A^{(3)}_v$-type initial states that belong to the same $O(h)$ class. Although the time dependence of the mean-square distance $d(r_0,t)$ for both states exhibits qualitatively similar behavior (Fig.\ref{Figure4}(c)), the small-$g$ behavior of the plateau residence time $T_p$ differs drastically between them. In one instance (Fig.\ref{fig:plateau-res-time-scaling}(d)), $T_p$ decreases monotonically with increasing $g$, consistent with the expectation that lowering the potential barrier enhances tunneling and hence reduces $T_p$. Remarkably, for the wavepacket dynamics shown in Fig.\ref{fig:plateau-res-time-scaling}(a), we instead find an anomalous trend in which $T_p$ grows with increasing $g$ in the small-$g$ regime.} \textcolor{blue}{On the other hand, upon approaching the weakly anisotropic regime ($g\rightarrow 1$), we find that $T_p$ decreases rapidly, independent of the particular choice of site-localized initial state. For perturbative values of $h/|\lambda_A^{(2)}|$, the plateau residence time exhibits a power-law scaling, $T_p\sim (1-g)^{\alpha_p}$, where the exponent $\alpha_p$ depends on the choice of the initial state but shows negligible dependence on the field strength \cite{comment-alphap-h-dependence}. For the two wavepacket dynamics considered here, we obtain $\alpha_p\approx 1.9$ (Fig.\ref{fig:plateau-res-time-scaling}(c)) and $\alpha_p\approx 1.0$ (Fig.\ref{fig:plateau-res-time-scaling}(d)). }

\begin{figure}
\includegraphics[width=1\columnwidth]{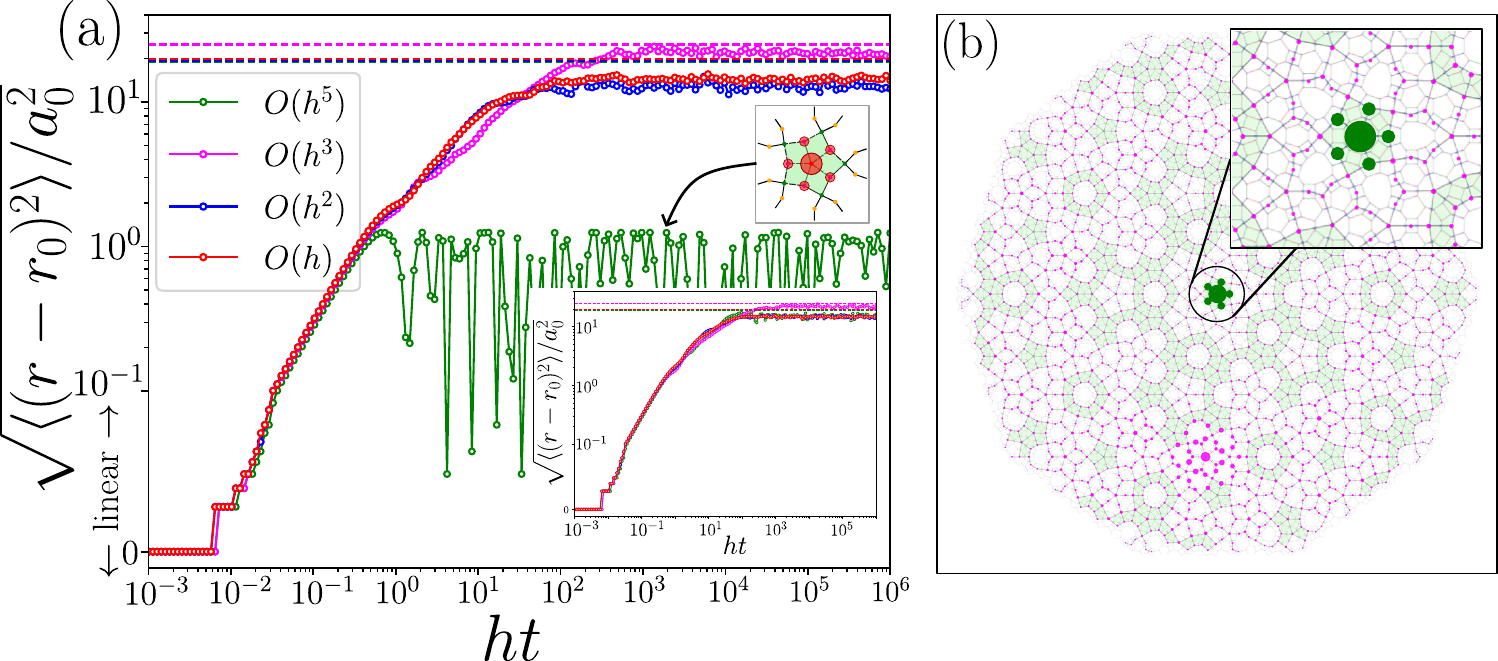}
\caption{{\bf $\pi$-flux localization of an $e$ anyon:~} \textcolor{blue}{When the $A_{v}^{(n)}$ couplings ($\lambda_{A}^{(n)}$) are uniform}, all $e$ anyon wavepackets delocalize except those initially placed on an $O(h^{5})$ site surrounded by an odd number of $\pi$-flux plaquettes (see Panel~(a) inset with an arrow). Such a wavepacket oscillates between the initial $A^{(5)}$ site and its five neighboring $A^{(3)}$ sites over exponentially long time scales, as shown by the green curve in Panel~(a) and the corresponding time-averaged probability density in Panel~(b). In contrast, an $O(h^{3})$ site—which exhibited localization in the presence of inhomogeneous $\lambda_{A}^{(n)}$ couplings (see Fig.\ref{Figure4})—now becomes fully delocalized, as seen in its mean-displacement curve Panel~(a) and $t$-averaged probability density Panel~(b). When the $\pi$-fluxes are removed, even the $O(h^{5})$ sites delocalize, as illustrated in the inset of Panel~(a). All the figures here are for Gen-7.}
\label{Figure5}
\end{figure}

\textcolor{blue}{We next analyze the scaling of $T_p$ with $h/|\lambda_{A}^{(2)}|$ for different values of $g$ close to unity. The Zeeman field range is chosen such that the excitations remain gapped throughout. Near $g=1$, we observe a power-law scaling, $T_p\sim h^{-\beta_p}$, with exponents that are roughly independent of $g$. For the wheel-like wavepacket dynamics (Fig.\ref{fig:plateau-res-time-scaling}(a)), we find $\beta_p\approx 3.0$, whereas for the arc-like dynamics (Fig.\ref{fig:plateau-res-time-scaling}(b)), we obtain $\beta_p\approx 2.0$.}

\textcolor{blue}{Interestingly, the observed scaling of $T_p$ with $h$ and $(1-g)$ deviates significantly from the naive perturbative expectation. From perturbation theory, one would anticipate $T_p^{-1}\sim h^{m}/(1-g)^{m-1}$, where $m$ denotes the minimum number of links required to connect two degenerate contours. In both cases studied here, $m\geq 4$. In the supplementary section \cite{plateau-time-note}, we further study a toy model of coupled rings designed to mimic the wavepacket dynamics shown in Fig.\ref{fig:plateau-res-time-scaling}(a). Unlike the TC quasicrystal, the toy model exhibits $h$- and $g$-dependent scaling consistent with perturbative expectations. These findings therefore indicate that, beyond the tunneling physics associated with the potential barrier, geometric constraints of the quasicrystal and the background-flux-induced interference effects can play a crucial role in generating the nontrivial scaling behavior of the wavepacket dynamics.}

\begin{figure*}
\includegraphics[width=1.0\textwidth]{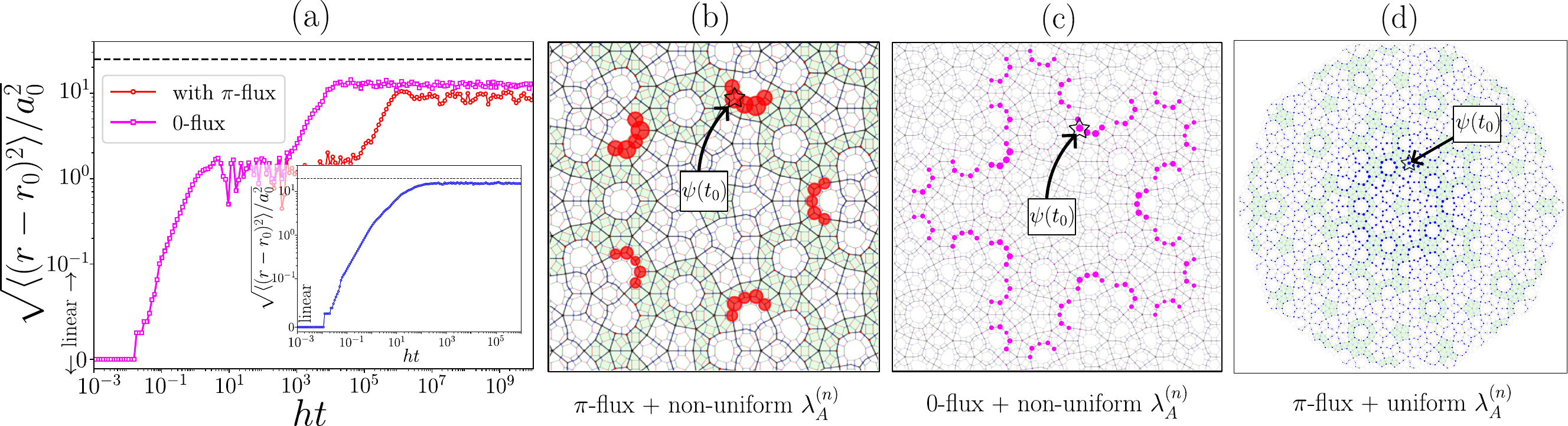}
\caption{{\bf Combined localization effect of $\pi$-flux and non-uniformity in $A_{v}^{(n)}$ couplings on wavepacket spreading:~} For a wavepacket initially localized at an $A^{(3)}$-type site with a hopping distance of $O(h)$, the presence of a background $\pi$-flux distribution helps suppress the spreading of the wavefunction. As shown in Panel~(a), the $\pi$-flux background delays the onset of the second plateau transition by a timescale of $O(10^{3})$. Furthermore, the time-averaged probability density in the flux-free case exhibits enhanced spreading [Panel~(c)] compared to the case with a $\pi$-flux background [Panel~(b)]. This localization tendency arises from the combined influence of the flux background and the spatial non-uniformity in the $A_{v}^{(n)}$ couplings ($\lambda_{v}^{(n)}$) experienced by the $e$ anyons. As illustrated in Panel~(d), in the absence of any non-uniformity $\lambda_{A}$'s, the flux background alone is insufficient to suppress the wavepacket spreading, and the average displacement ($d(r_{0},t)$) quickly approaches the delocalized limit within a much shorter timescale, $ht \sim 10^{2}$ [Panel~(a), inset].}
\label{Figure6}
\end{figure*}    

\section{$\pi$-Flux Localization}
\label{sec:piflux}

Interestingly, the structure of the quasicrystalline TC lattice itself can cause the background $\pi$ flux (static $m$ anyon) configuration to strongly confine the $e$ anyons on certain lattice sites, even when the onsite energies ($\lambda_{A}^{(n)}$) are completely uniform. While the small-$g$ expansion of the underlying Kitaev model breaks down in this regime, we nevertheless analyze this limit to explore how quantum interference effects of the $\pi$ fluxes can influence the motion of $e$ anyons. 

The star-shaped cluster (see Fig.\ref{Figure3-part1}(a)) associated with this flux-induced localization features an $A^{(5)}$-type vertex at its center and is surrounded by five $\pi$-flux plaquettes (collectively representing a single $m$ excitation). Such motifs typically appear at the center of odd-generation lattices, although they can also emerge elsewhere across both even and odd generations. We compute the mean-square displacement $d(r_{0},t)$ for the same set of $A^{(5)}$-type starting vertices ($r_{0}$) used in Fig.\ref{Figure4}(a) and (b), now imposing uniform onsite potentials ($\lambda_{v}^{(n)}=\lambda$ for all $v$). Aside from the $O(h^{5})$ vertex (green), which lies within this star cluster, the wavepackets initialized at other vertices disperse quickly to equipotential sites, and $d(r_{0},t)$ reaches its delocalized value on a timescale $ht\sim 10^{2}$—significantly shorter than in the non-uniform potential case (Fig.\ref{Figure5}). Moreover, this delocalization time is nearly independent of $r_{0}$. In contrast, a wavepacket starting at the star cluster exhibits oscillatory motion between the central $A^{(5)}$ site and its five surrounding $A^{(3)}$ neighbors (see Fig.\ref{Figure5}). This confinement stems purely from interference due to the $\pi$ flux, which can be verified in multiple ways: (1) An $O(h^{3})$ vertex not encircled by any $\pi$-flux (or any odd number of them) was completely localized under non-uniform potentials (Fig.\ref{Figure4}(a)), yet once the potential variation is removed, the wavepacket spreads across the entire lattice (Fig.\ref{Figure5}(b)). (2) In the zero-flux case (Fig.\ref{Figure5}(a) inset), the $O(h^{5})$ vertex delocalizes on the same timescale as all other vertices. These findings unambiguously demonstrate the central role of $\pi$ flux in producing wavepacket localization. Most notably, there exists a finite-energy eigenstate of $H_{b}$ (with flux background) that possesses nodes at the five outer corner sites of the star cluster (see SM), effectively isolating this cluster from the remainder of the lattice and resulting in strong localization.

The composite excitation formed by an $e$ and an $m$ anyon—referred to as the $\varepsilon$ particle—exhibits fermionic exchange statistics. In the square-lattice TC model, such bound states can only be generated in pairs through the transverse Zeeman coupling, $\sum_{l}\tau^{y}_{l}$, and these fermionic excitations display constrained, fracton-like mobility in lower dimensions \cite{PhysRevB.80.081104, PhysRevB.108.035149}. In the quasicrystalline TC model with uniform couplings, however, the emergence of an $e$–$m$ bound state arises from the geometric structure of the lattice, which causes the $\pi$ fluxes to trap the $e$ anyon. This mechanism differs from scenarios where $\varepsilon$ fermions appear as zero modes bound to topological defects \cite{PhysRevB.87.045106, PhysRevB.90.134404}; here, the lattice geometry alone yields a unique route to realizing stable $\varepsilon$ excitations at finite energies.

Apart from complete localization, the background flux also amplifies the suppression of wavepacket spreading that arises from the non-uniform potential landscape. As shown in Fig.\ref{Figure6}(a), the mean-square displacement $d(r_{0},t)$ for a representative $A^{(3)}$-type vertex reaches the second plateau at a substantially later time when the background $\pi$-flux distribution is included alongside the potential variations (red curve), compared to the corresponding zero-flux scenario (magenta curve). The associated delay is of order $O(10^{3})$ in units of $h$. Moreover, the spatial extent over which the wavepacket spreads is larger in the absence of flux, which becomes evident by comparing the time-averaged probability distribution at the second plateau for the $\pi$-flux case (Fig.\ref{Figure6}(b)) with that of the 0-flux case (Fig.\ref{Figure6}(c)). This slowdown of delocalization results from the combined effects of flux-induced interference and tunneling across the potential barriers. As evident from the inset of Fig.\ref{Figure6}(a) and Fig.\ref{Figure6}(d), the flux configuration by itself is insufficient to prevent delocalization, since $d(t)$ approaches its delocalized saturation value over a significantly shorter timescale ($ht \sim 10^{2}$) when the onsite potentials are uniform.

\section{Role of phason fluctuations}
\label{sec:phason}
\textcolor{blue}{Phasons correspond to novel collective degrees of freedom, which are completely inherent to the quasicrystals \cite{PhysRevB.34.3345, PhysRevLett.134.136003}. A phason fluctuation is a smooth, volume-preserving deformation of the projection window in the perpendicular space of the higher-dimensional periodic lattice from which the quasicrystal is generated. In the physical space, such fluctuations manifest as local rearrangements of atoms, bonds, or tiles within the quasicrystalline structure. These rearrangements locally perturb the quasiperiodic order and can therefore influence both the low-energy collective phases and the dynamics of quasiparticles in nontrivial ways. While the effects of phason disorder on the elastic and electronic properties of quasicrystals have been explored extensively \cite{widom2008discussion, de2008phason}, comparatively little attention has been devoted to its impact on quantum magnetism \cite{phason-qc-afm, PhysRevB.85.094434, goldman2014magnetism}, and studies addressing its role in topologically ordered phases \cite{topology-quasicrystal-prx} remain even more limited.} 

\begin{figure}
    \centering
    \includegraphics[width=1.0\columnwidth]{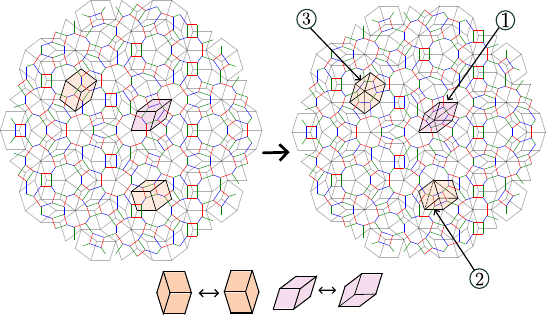}
    \caption{Phason-flips in the Penrose quasicrystal and the corresponding plaquette rearrangements in the Kitaev quasicrystal. Two different types of local rearrangements of the rhombic tilings (light orange and light violet) create three distinct types of elementary phason defects in the Kitaev lattice.}
    \label{fig:phason-flip-penrose-qc}
\end{figure}

\begin{figure}
    \centering
    \includegraphics[width=1.0\columnwidth]{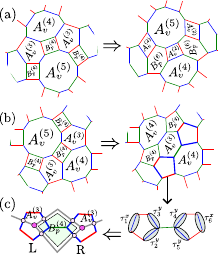}
    \caption{Effect of phason flips on the Kitaev quasicrystal and the resulting changes of the TC stabilizer operators. (a) In the presence of a type-2 phason defect (Fig.\ref{fig:phason-flip-penrose-qc}), the local plaquette configuration is altered; however, the tri-coordinated coloring structure of the Kitaev lattice remains intact. (b) A type-3 phason defect generates new pentagonal plaquettes, leading to a breakdown of the tricoordinated coloring structure. (c) TC stabilizer for the pentagon-dimer configuration and its decomposition into a product of $A^{(3)}_v$ and $B_{p}^{(4)}$ operators.} 
    \label{fig:phason-flip-tc-ops}
\end{figure}

\textcolor{blue}{Phason flips can be generated in the quasicrystalline Kitaev model through two inequivalent ways. One possibility is to generate these defects by rearranging the rhombic tiles of the Penrose quasicrystal, which subsequently induces local modifications in the plaquette structure of the associated tri-coordinated Kitaev lattice. Alternatively, phason flips may be introduced directly on the tri-coordinated quasicrystal itself. Here, we take the former approach, as the local geometric characteristics of phason flips in the Penrose tiling are well established. Fig.\ref{fig:phason-flip-penrose-qc} illustrates two representative phason-flip processes on the parent Penrose quasicrystal together with the corresponding changes in the plaquette structure of the Kitaev lattice. We identify three distinct classes of local plaquette rearrangements (Fig.\ref{fig:phason-flip-penrose-qc}). The type-1 phason flip leaves the local plaquette topology unchanged, producing only distortions in the bond angles. By contrast, type-2 and type-3 involve nontrivial modifications of the local plaquette structure (see Fig.\ref{fig:phason-flip-tc-ops}(a) and (b)). In particular, type-3 leads to the formation of new pentagonal plaquettes and simultaneously disrupts the bond coloring pattern of the tri-coordinated parent lattice (Fig.\ref{fig:phason-flip-tc-ops}(b)). As a result, a site may have more than one bond carrying the same color label. In the Hamiltonian description, this implies that identical easy-axis Ising interactions can appear on multiple bonds connected to the same site.}

\textcolor{blue}{The breakdown of the one-color-per-link condition of the tri-coordinated lattice renders the resulting Kitaev Hamiltonian non-integrable, as the $\mathbb{Z}_2$ flux operators are no longer well defined on odd-sided plaquettes. Consequently, phason fluctuations generically transform the system into an interacting Majorana problem for arbitrary values of $J_{\alpha}$, with random nearest-neighbor two-body interactions among the Majorana fermions. Nevertheless, one can still consider the TC limit in the spin language, as we illustrate in Fig.\ref{fig:phason-flip-tc-ops}. For type-1 and type-2 phason defects, no new stabilizer operators are generated; the defects merely modify the local arrangement of the existing $A_v$ and $B_p$ operators. The situation is different for type-3 defects. In this case, besides creating a local rearrangement of the conventional QCTC stabilizers, the defect gives rise to a new mixed-type stabilizer operator in the low-energy Hamiltonian, 
\begin{align}
\sim -\frac{9J^5}{64J^4_{z}}\tau^x_{1}\prod_{i=2}^5\tau^y_{i}\tau^x_6 = -\frac{9J^5}{64J^4_{z}} A^{(3)}_{v,L} B^{(4)}_{p}A^{(3)}_{v,R}
\end{align}
Since this is the only leading-order term for the pentagon-dimer plaquette, in the ground state, the above operator should take $+1$ expectation value. The corresponding eigenstate ($J>0$) is
\begin{align}
\ket{\psi_{\text{mix}}}=\frac{1}{2}(\ket{0}+\ket{e_L,e_R}+\ket{e_{L}, m}+\ket{e_R, m})
\end{align}
which is an equal amplitude linear superposition of trivial ($\mathbb{Z}_2$ charge-free) and topological \textit{dyonic} states. In Secs.~\ref{sec:piflux} and \ref{sec:dyne}, we showed that geometric constraints and potential scattering can localize the $e$-$m$ bound state near the center of the quasicrystal. Here, we uncover a different phenomenon: a particular class of lattice defects can stabilize a fractional density of fermionic quasiparticles already in the ground state. While the absence of a global $U(1)$ symmetry precludes a well-defined notion of charge in this system, these results raise intriguing questions regarding the nature of the thermodynamic phases that emerge in the presence of a finite Zeeman field. In particular, it would be worthwhile to investigate how these defect-induced fermionic excitations interact with the other Abelian anyonic quasiparticles and influence the low-energy phases of the system.}
\section{Conclusion}
\label{sec:conc}
In this work, we derived the effective low-energy TC Hamiltonian that emerges in the large-$J_{z}$ limit of the exactly solvable tri-coordinated QC Kitaev spin liquid. Owing to the presence of \textcolor{blue}{plaquettes of multiple sizes} in the QC lattice, the effective TC theory hosts star and plaquette stabilizers of different orders, with coupling strengths separated over exponentially large scales. Remarkably, the resulting exactly solvable TC ground state contains a finite (irrational number) density of both $e$ and 
$m$ anyons, even in the absence of external perturbations. The interplay between this background anyon density and the strong anisotropy in the coupling parameters gives rise to anomalous localization behavior in the low-energy excitations. In Sec.~\ref{sec:groundst}, and Sec.\ref{sec:zeeman}, we described the exact ground-state structure and examined how excitations evolve under Zeeman fields. Depending on their geometric origin, some excitations acquire linear dispersion ($\propto h$), whereas others disperse quadratically ($\propto h^{2}$). We further showed that localized and delocalized eigenstates can coexist at the same energy, a consequence of the simultaneous energetic and geometric isolation of states on the QC lattice. In Sec.~\ref{sec:dyne}, we analyzed the dynamics of $e$ charges and uncovered strongly anomalous localization phenomena: depending on the initial condition, a wavepacket may remain strictly localized or undergo sequential delocalization across contours of increasing size. While these quasilocalized dynamics primarily stem from the strong coupling anisotropy, Sec.~\ref{sec:piflux} demonstrates that regions of $\pi$-flux (or static $m$ charges), together with geometric constraints, can also generate highly localized states. \textcolor{blue}{Although a quantitatively reliable characterization of localization in terms of localization lengths or diffusion exponents requires a systematic finite-size scaling analysis over sufficiently large system sizes, we expect several qualitative features, such as plateau-like behavior in the wavepacket dynamics, coexistence of localized and extended states at the same energy, and the emergence of $\pi$-flux-induced bound states to remain robust in the thermodynamic limit. This robustness arises because these phenomena are primarily governed by the local geometric constraints imposed by the quasicrystalline tiling pattern, rather than by the overall system size.} \textcolor{blue}{In Sec.\ref{sec:phason}, we studied the effects of phason fluctuations in the TC quasicrystal and showed how phason flips induce local rearrangements of the TC plaquettes, leading to the emergence of new mixed-type stabilizer operators that can stabilize nontrivial topological charges in the ground state. Our analysis, however, focused primarily on isolated phason defects. An important direction for future work would be to investigate how the ground-state properties evolve in the presence of a finite density of such defects, particularly when the phason degrees of freedom acquire their own dynamics. Another promising direction is to study the transport and diffusion properties of Abelian anyons in the presence of thermally fluctuating phason disorder \cite{phason-defect-electron, Phason-vidal}, where the interplay between topological excitations and dynamical geometric disorder may give rise to novel phenomena.}

While our analysis has focused on the TC phase in a specific quasicrystalline geometry, it will be equally interesting to look at this phase in other quasicrystalline \cite{PhysRevB.108.104208} and hyperbolic \cite{mx1t-74dm} geometries, and especially in the amorphous setting to see how the chiral spin liquid transits to the gapped TC phase \cite{cassella2023exact, KimPRL2023}. 

Although we have qualitatively argued for the nature of the fractionalized phase beyond perturbative Zeeman couplings and sketched the broad structure of the phase diagram (see \Fig{Figure2}), a complete quantitative understanding of the full phase diagram will require more computationally intensive methods, particularly to uncover how confinement physics develops in this setting\cite{PhysRevB.82.085114, PhysRevLett.117.210401, PhysRevX.11.041008, 3lxj-dx76, PhysRevB.105.075132, PhysRevResearch.6.013298}. Even at the perturbative-field level, the anyon dynamics already exhibits features reminiscent of a Bose-glass \cite{yu2012bose}, and it would be valuable to obtain a more quantitative characterization of such glassy phases within the context of topologically ordered $\mathbb{Z}_{2}$ spin-liquids \cite{PhysRevLett.114.247207, PhysRevLett.118.087203, zhu2025emergent, PhysRevB.108.165118, hart2021correlation}. Similar glassy behavior has been reported previously in disordered TC \cite{PhysRevB.83.075124}. In translationally invariant systems, the field-induced confinement transition is typically understood as arising from the condensation of bosonic $e$ and $m$ charges. Here, however, geometric constraints prevent the formation of system-spanning delocalized states, raising the possibility that the confinement transition may not be second order. Moreover, the quasi-periodic distribution of background 
$\pi$-fluxes and the discrete rotational symmetries of the quasicrystal may play a key role near the transition—for instance, the critical theory (if the transition is second-order) could exhibit emergent symmetries \cite{borla2025oddtoriccode} not apparent in the $\mathbb{Z}_{2}$ symmetric microscopic model.

Another interesting direction for future study concerns the role of dynamical electronic charge degrees of freedom in such spin-liquid phases. The impact of hole doping on spin liquids has been a long-standing topic of interest since the early investigations of high-$T_{c}$ superconductors \cite{PhysRevLett.61.2376, RevModPhys.78.17}, and more recently in the context of topological superconductivity in Kitaev spin liquids \cite{PhysRevB.86.085145, PhysRevB.85.140510, PhysRevB.109.184508, hardy2025hunting} in translationally invariant lattices. In quasicrystals, the nontrivial (and often fragmented) single-particle band structure, as well as the fractal-like density of states generated by long-range aperiodic order, give rise to highly inhomogeneous and pattern-selective superconducting pairing tendencies \cite{6b35-y4zj, biswas2025,PhysRevLett.134.206001}, and can, in certain regimes, even enhance the stability of topological superconductivity \cite{PhysRevB.110.134508}. It would be interesting to understand how these phenomena are modified in the strongly correlated regime.

With the emergence of new experimental platforms capable of realizing quasicrystalline systems, addressing these questions will be an exciting direction for future work.

\section{Acknowledgments}  
We acknowledge fruitful discussions with Subrata Pachhal. SS acknowledges support from Institute Postdoctoral Fellowship, IIT Kanpur. AA and MS acknowledge prior collaboration with Sunghoon Kim, Animesh Nanda, Debanjan Chowdhury and Subhro Bhattacharjee on related themes. We acknowledge Julien Vidal for discussions, critical reading of the manuscript and pointing out an error in an earlier version. MS also acknowledges financial support from Indian Institute of Technology Kanpur as well as from the Department of Physics, University of Illinois Urbana-Champaign. AA acknowledges support from IITK Grants (IITK/PHY/2022010) and (IITK/PHY/2022011). Numerical calculations were performed on the workstations {\it Wigner} and {\it Syahi} at IITK. 

\bibliography{refs_QC_TC}

\newpage
\setcounter{section}{0} 
\setcounter{equation}{0}
\makeatletter
\renewcommand{\theequation}{S\arabic{equation}}
\setcounter{figure}{0}
\renewcommand{\thefigure}{S\arabic{figure}}

\onecolumngrid

\begin{center}
	\textbf{\large Supplemental Material for ``Anyon Quasilocalization in a Quasicrystalline Toric Code''}
\end{center}

\vspace{\columnsep}
\vspace{\columnsep}



\section{Determining the fractions of different plaquettes of the toric code lattice}

\begin{figure}
    \centering
    \includegraphics[width=0.7\linewidth]{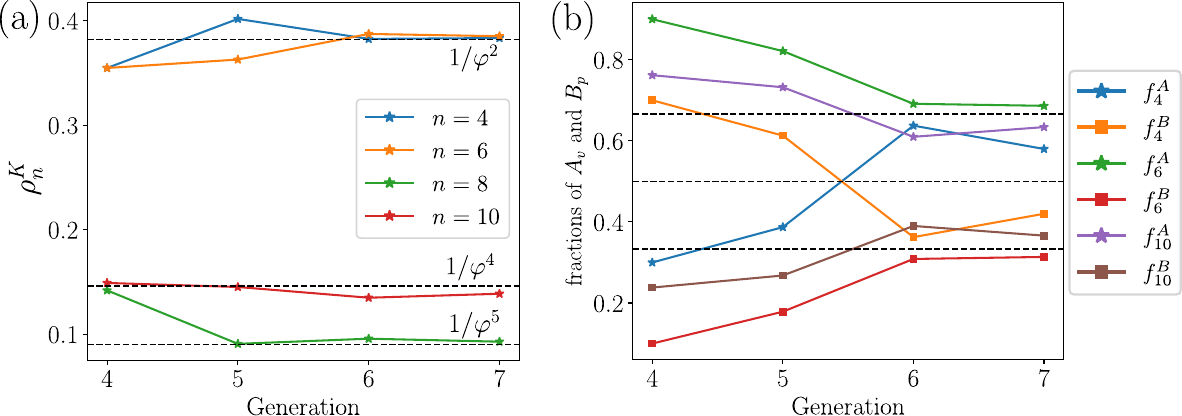}
    \caption{{\bf Densities of star and plaquette operators}: (a) Fraction (or density) of plaquettes $\rho_{n}^{K}$ having $n$ sides of the tri-coordinated Kitaev quasicrystal for different generations. (b) For a given plaquette size $n$, the fractions of $A_{v}$-type plaquettes ($f^{A}_{n}$) and  $B_{p}$-type plaquettes ($f^{B}_{n}$) is plotted for different generations.}
    \label{fig:plaquette-fractions}
\end{figure}

The densities of various plaquette types in the tri-coordinated Kitaev quasicrystal were previously determined in Ref.~\cite{PhysRevB.110.214438}. The fractions ($\rho_{n}^{K}$) of square ($n=4$) and hexagonal ($n=6$) plaquettes converge rapidly to $\varphi^{-2}$ with increasing generation number (or system size), while those of decagonal ($n=10$) and octagonal ($n=8$) plaquettes approach $\varphi^{-4}$ and $\varphi^{-5}$, respectively. In Fig.~\ref{fig:plaquette-fractions}(a), we reproduce these results for completeness.

A plaquette of size $n$ in the tri-coordinated lattice can give rise to either a star ($A_{v}$) or a flux ($B_{p}$) term in the corresponding toric-code lattice. Accordingly, we define two fractions as
\begin{align}
&f_{n}^{A} = \frac{\text{number of plaquettes of size } n \text{ generating } A_{v} \text{ terms}}{\text{total number of plaquettes of size } n}, \\
&f_{n}^{B} = \frac{\text{number of plaquettes of size } n \text{ generating } B_{p} \text{ terms}}{\text{total number of plaquettes of size } n}.
\end{align}
As shown in Fig.~\ref{fig:plaquette-fractions}(b), the fractions $f_{6}^{A}$ and $f_{6}^{B}$ converge clearly to $2/3$ and $1/3$, respectively. Similarly, $f_{10}^{A}$ and $f_{10}^{B}$ approach the same ratios, though with slower convergence. The fractions for $n=4$ show oscillatory behavior around $1/2$; however, due to finite-size limitations, we cannot conclusively determine whether they really converge to $1/2$ in the thermodynamic limit.

This behavior can be understood from the structural property of the original quasicrystalline lattice: none of its plaquettes contains all three bond types ($xx$, $yy$, and $zz$) along their edges. The square plaquettes appear in two varieties—(1) those composed of $xx$ and $yy$ bonds, and (2) those composed of $xx$ and $zz$ bonds. Hexagonal plaquettes, on the other hand, occur in three types—(1) with $xx$ and $yy$ bonds, (2) with $xx$ and $zz$ bonds, and (3) with $yy$ and $zz$ bonds. The same classification applies to the decagonal plaquettes. Assuming that in the thermodynamic limit these different plaquette types occur with equal probability, one naturally obtains the above fractions.

\section{Derivation of quasicrystalline toric code model}
\begin{figure}
    \centering
    \includegraphics[width=0.9\linewidth]{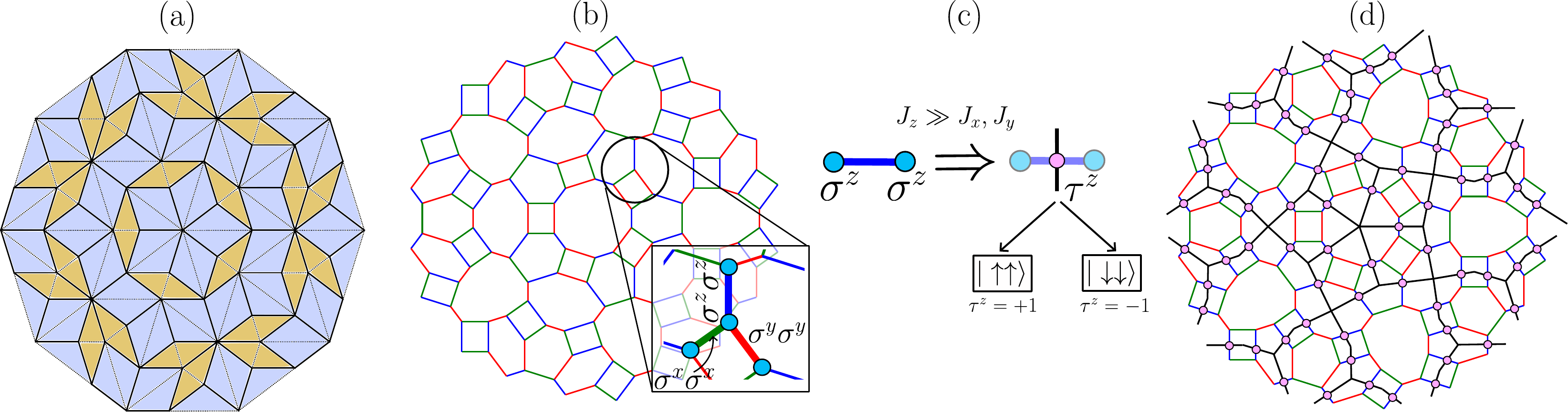}
    \caption{{\bf Toric code lattice construction:}
(a) The generation-3 (Gen-3) Penrose quasicrystal, composed of two types of tiles—the golden triangle and the golden gnomon. (b) By connecting the centroids of these triangles, one obtains the Gen-3 tri-coordinated Kitaev quasicrystal. Each vertex carries three bonds corresponding to bond-dependent Ising interactions: Green denotes $\sigma^{x}\sigma^{x}$, Red denotes $\sigma^{y}\sigma^{y}$, and Blue denotes $\sigma^{z}\sigma^{z}$, with coupling strengths $J_{\alpha}$ ($\alpha = x, y, z$).
(c) In the large-$J_{z}$ limit, an effective bond pseudo-spin degree of freedom ($\tau$) emerges on each $z$-bond.
(d) The resulting dual toric-code lattice, on which these bond pseudo-spins reside at the centers of the links of the dual graph.}
    \label{fig:TC-construction}
\end{figure}

The effective low-energy Hamiltonian in the limit of $J_{z}\gg J_{x}, J_{y}$ can be derived using time-independent degenerate perturbation theory in the small parameter, $J_{\alpha}/J_{z}$ ($\alpha=x,y$). The zeroth order Hamiltonian is given by,
\begin{align}
H_{0}=-J_{z}\sum_{z\text{-bonds}}\sigma^{z}_{i}\sigma^{z}_{j} 
\end{align}
The ground state (GS) subspace of each $z$-link is two-fold degenerate, $\ket{GS}=\lbrace \ket{\uparrow\uparrow}, \ket{\downarrow\downarrow}\rbrace$, which leads to an effective pseudospin at every $z$-link. The dual toric code (TC) lattice can be constructed by joining links that intersect the strong $z$-bonds of the tri-coordinated Kitaev lattice, and the $\tau$ spins are placed on these bonds (see Fig.\ref{fig:TC-construction} for the dual lattice construction). The tri-coordinated Kitaev lattice and its dual TC lattice do not contain the same number of sites. Table \ref{tab:gen-sites} lists the site counts for both lattices across four successive generations of increasing system size.

\begin{table}
    \centering
    \begin{tabular}{c|c|c}
         \textbf{Gen} & \textbf{No. of sites} & \textbf{No. of sites}\\
          & \textbf{(Kitaev lattice)} & \textbf{(toric code lattice)}\\
          \hline
         4 &  340 & 86  \\
         5 &  890 & 251 \\
         6 &  2330 & 721 \\
         7 &  6100 & 1906
    \end{tabular}
    \caption{Number of lattice sites in the tri-coordinated Kitaev lattice (second column) and in the dual toric-code lattice (last column) for various lattice generations (Gens).}
    \label{tab:gen-sites}
\end{table}

We define the Pauli pseudospin operators, $\tau^{x}$ and $\tau^{z}$ which only act within the GS subspace,
\begin{align}
\tau^{z}=\ket{\uparrow\uparrow}\bra{\uparrow\uparrow}-\ket{\downarrow\downarrow}\bra{\downarrow\downarrow}\\
\tau^{x}=\ket{\uparrow\uparrow}\bra{\downarrow\downarrow}+\ket{\downarrow\downarrow}\bra{\uparrow\uparrow}
\end{align}  

We are interested in projecting the following perturbation Hamiltonian
\begin{align}
V=-J_{x}\sum_{x\text{-bonds}}\sigma^{x}_{i}\sigma^{x}_{j}-J_{y}\sum_{y\text{-bonds}}\sigma^{y}_{i}\sigma^{y}_{j}
\end{align}
to the ground state subspace. This projection will result in an effective Hamiltonian $H_{\text{eff}}$ that couples the bond pseudospins via various multi-spin interactions. The general form of $H_{\text{eff}}$ can be written as a series expansion in $V$ \cite{kitaev2006anyons},
\begin{align}
H_{\text{eff}} = H_{\text{eff}}^{(1)} + H_{\text{eff}}^{(2)} + H_{\text{eff}}^{(3)} + \cdot \cdot \cdot \cdot
\end{align}
where $H_{\text{eff}}^{(1)} = P V P$ and 
\begin{align}
H_{\text{eff}}^{(n)} = P \prod_{j=1}^{n-1} \bigg[V(1-P)\frac{1}{E_{0}-H_{0}}\bigg](1-P)VP
\end{align}
A single application of $V$ simultaneously excites two $z$ bonds. To return to the ground-state subspace, $V$ must either be applied again on the same bond (which merely renormalizes the onsite energy) or act successively along a closed loop. Consequently, contributions from larger loops appear only at higher orders in the perturbation expansion. \\

\noindent The plaquettes (or loops) of the original tri-coordinated quasicrystal can be grouped into two different categories.\\

\begin{figure}
    \centering
    \includegraphics[width=0.5\linewidth]{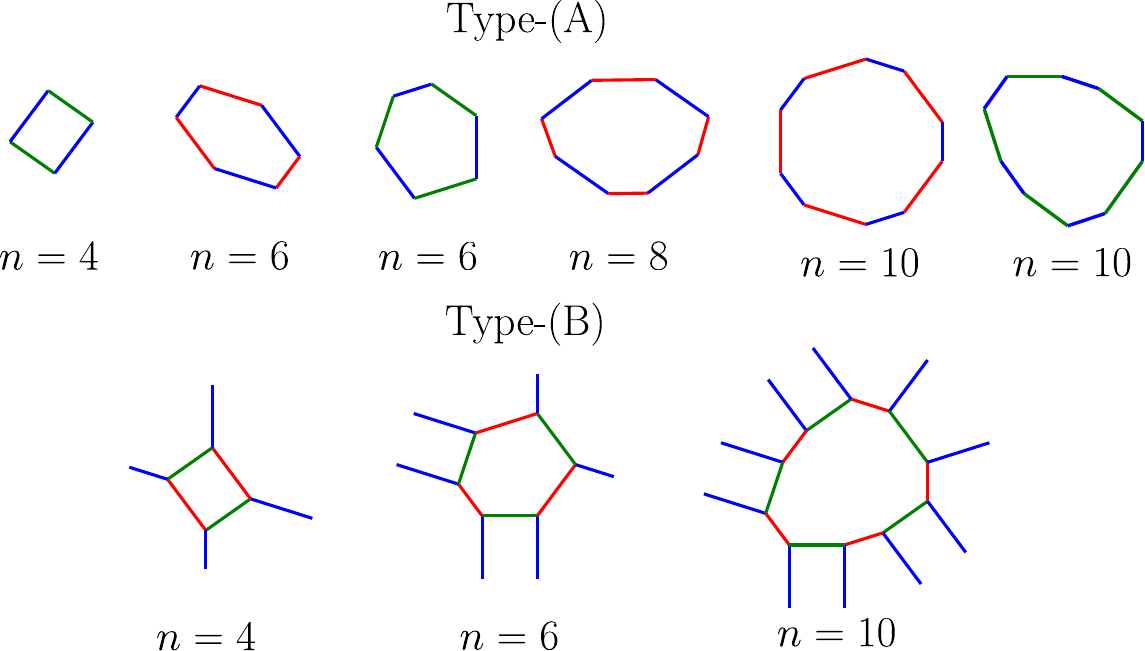}
    \caption{{\bf Different types of minimal plaquettes of the tri-coordinated quasicrsytal}: Different types of plaquettes of the tri-coordinated quasicrystal that generates different multi-spin operators ($A_{v}$ and $B_{p}$), under strong-coupling perturbation theory.}
    \label{fig:TC-plaquettes}
\end{figure}

\noindent Type-(A): In this case, the $z$-bonds form part of the plaquette edges. In such cases, the $z$-bonds are connected through either $x$ or $y$ links (but not both simultaneously). These loops give rise to star operators ($A_{v}^{(n)}$) of various orders (Fig.\ref{fig:TC-plaquettes}).

\noindent Type-(B): For some plaquettes, the $z$-bonds (blue links) emanate from the plaquette boundaries. These plaquette boundaries contain only $x$ (green) and $y$ (red) bonds. There are three such plaquettes which will generate the $B^{(n)}_p$ terms (Fig.\ref{fig:TC-plaquettes}).

As will be illustrated through explicit examples, in the case of type-(B) plaquettes, the sequence of successive perturbations affects both the numerator (through the transition matrix elements) and the denominator (through the energies of the intermediate states) of the perturbation series. In contrast, for type-(A) plaquettes, only the energy denominators are sensitive to the ordering of the perturbation operators.\\

\noindent {\textit{Effective Hamiltonian at different orders}}: First order correction is zero because a single action of $V$ always leads to an excited state configuration.

Second order: In contrast to Kitaev's toric code, the quasicrystalline model has non-trivial terms even at the second order ($\sim A_{s}^{(2)}$) that arise from the plaquettes (of the original lattice) having only the z and x bonds. There are only two equivalent ways of applying the $xx$ perturbation, and both of these lead to the same energy cost $4J_{z}$ (energy required to break two $z$-bonds). Hence, we obtain,
\begin{align}
H_{\text{eff}}^{(2)}=2\times \frac{(-J_{x})^{2}}{(-4J_{z})}\tau^{x}_{1}\tau^{x}_{2}\equiv -\frac{J_{x}^{2}}{2J_{z}}A_{v}^{(2)}    
\end{align}
There are no such plaquette ($B_{p}$) terms at this order of expansion.

Third order: There are two contributions: (1) from the hexagonal plaquettes (of the tri-coordinated lattice) that consists of only z and x bonds, and (2) the hexagonal plaquettes having only z and y bonds as their sides. There is only one sequence of the intermediate excitation energies, $0\rightarrow 4\rightarrow 4\rightarrow 0$ (in the units of $J_{z}$) and all 3! possible orderings (of $V$) are equivalent. Thus, for type (1) hexagon, we get
\begin{align}
H_{\text{eff},1}^{(3)}=3!\times\frac{(-J_{x})^{3}}{(-4J_{z})^{2}}\prod_{j=1}^{3}\tau^{x}_{j}\equiv -\frac{3J_{x}^{3}}{8J_{z}^{2}}A_{v}^{(3)}
\end{align} 
For the type (2) hexagonal plaquette, there is an overall multiplicative factor $(\pm i)^{6}$ (since $\sigma^{y}\ket{\pm}_{z}=(\pm i)\ket{\mp}_{z}$). Therefore, in this case, an extra minus sign appears,
\begin{align}
H_{\text{eff},2}^{(3)}=+\frac{3J_{y}^{3}}{8J_{z}^{2}}A_{v}^{(3)}
\end{align}
Again, there are no $B_{p}^{(3)}$ terms. 

Fourth order: In this order of expansion, one $A_{v}$ and one $B_{p}$ term is generated. The $A_{v}$ ($B_{p}$) term arises from octagons (squares) having only $z$ and $y$ ($x$ and $y$) bonds at the boundaries. Calculation of $A_{v}^{(4)}$ follows exactly the same steps as done for $A_{v}^{(3)}$. There are $4!$ equivalent orderings of $V$ and two different sequence of intermediate energy denominators: ($0\rightarrow 4 \rightarrow 4 \rightarrow 4 \rightarrow 0$) and ($0\rightarrow 4 \rightarrow 8 \rightarrow 4 \rightarrow 0 $). After adding all of thenm, we obtain,
\begin{align}
H_{\text{eff},1}^{(4)}=-\bigg[\frac{16}{64}+\frac{8}{128}\bigg]\frac{J_{y}^{4}}{J_{z}^{3}}A_{v}^{(4)}=-\frac{5J^{4}_{y}}{16J^{3}_{z}}A^{(4)}_{v}\ \ \text{where},\ \ A_{v}^{(4)}=\prod_{j=1}^{4}\tau^{x}_{j}
\end{align}
Calculation of the $B_{p}^{(4)}$ proceeds in the following manner. There are 4 bond perturbations (two $xx$ and two $yy$). So, in total, we have 4! possible orderings of these perturbations. There are only two possible sequences of the intermediate energy denominators: ($0\rightarrow 4\rightarrow 4\rightarrow 4\rightarrow 0$) and ($0\rightarrow 4\rightarrow 8\rightarrow 4\rightarrow 0$); these terms add in the following manner,
\begin{align}
4\times\bigg[+\frac{2}{64}-\frac{2}{64}-\frac{2}{128}\bigg]\times \frac{J_{x}^{2}J_{y}^{2}}{J_{z}^{3}}
\end{align}
Therefore,
\begin{align}
H_{\text{eff},2}^{(4)}=-\frac{J_{x}^{2}J_{y}^{2}}{16 J_{z}^{3}} B_{p}^{(4)}\ \ \text{where}\ \ B_{p}^{(4)}=\prod_{j=1}^{4}\tau^{z}_{j}
\end{align}

Fifth order: In this order, $B_{p}$ terms do not arise, which is consistent with the time-reversal symmetry. A non-zero $A_{v}$ contribution arises from the decagonal plaquettes (of the tri-coordinated lattice). There are two types of them: (1) plaquettes that contain only $z$ and $y$ bonds, (2) plaquettes that contain only $z$ and $x$ bonds (see Fig.\ref{fig:TC-plaquettes}). There are $5!$ equivalent orderings of $yy$ and $xx$ perturbations. The possible sequences of the intermediate state energy denominators (in both cases) are: ($0\rightarrow 4\rightarrow 4\rightarrow 4 \rightarrow 4\rightarrow 0$), ($0\rightarrow 4\rightarrow 4\rightarrow 8\rightarrow 4\rightarrow 0$), ($0\rightarrow 4\rightarrow 8\rightarrow 4\rightarrow 4\rightarrow 0$), and ($0\rightarrow 4\rightarrow 8\rightarrow 8\rightarrow 4\rightarrow 0$). 
After collecting all of these processes, we obtain the following numerical pre-factor
\begin{align}
&\text{For plaquette type-(1): }\ \ \ \  5\times (i^{2})^{5}\times \frac{(-J_{y})^{5}}{J_{z}^{4}}\bigg[ \frac{8}{(-4)^{4}}+\frac{8}{(-4)^{2}(-8)^{2}}+\frac{8}{(-4)^{3}(-8)}\bigg]\\
&\text{For plaquette type-(2): }\ \ \ \  5\times \frac{(-J_{x})^{5}}{J_{z}^{4}}\bigg[ \frac{8}{(-4)^{4}}+\frac{8}{(-4)^{2}(-8)^{2}}+\frac{8}{(-4)^{3}(-8)}\bigg]
\end{align}
Therefore,
\begin{align}
H_{\text{eff},1}^{(5)}=+\frac{35}{128}\frac{J_{y}^{5}}{J_{z}^{4}}A_{v}^{(5)}\ \ ,\ \ H_{\text{eff},2}^{(5)}=-\frac{35}{128}\frac{J_{x}^{5}}{J_{z}^{4}}A_{v}^{(5)}\ \ , \text{where}\ A_{v}^{(5)}=\prod_{j=1}^{5}\tau^{x}_{j}
\end{align}
Sixth order: In this order of expansion, the effective Hamiltonian has only the $B_{p}^{(6)}$ terms that arises from the hexagonal plaquettes having $xx$ and $yy$ bonds as their sides. There are also so-called ``disconnected plaquette'' contributions that arise from the product of two lower-order (in $J_{\alpha}/J_{z}$) plaquette operators; for example, $\sim -\frac{J_{x}^{6}}{J_{z}^{5}} (A_{v}^{(3)})_{i}(A_{v}^{(3)})_{j}$. We will ignore these contributions since the dominant lower-order terms, $-(J_{x}^{3}/J_{z}^{2})A_{v}^{(3)}$, for the example considered, already decide the ground state flux configurations. Neglecting such ``self-energy'' terms should not affect the leading-order description. 

To calculate $B_{p}^{(6)}$, we have to consider all possible permutations of the 6 bonds. There are $6!$ orderings of the $xx$ and $yy$ bonds and these $6!$ terms can be grouped according to the sequences of the intermediate energies (see  
table (\ref{tab:table1})). For a given energy sequence, some terms lead to an overall positive or negative sign for the numerator. These are determined numerically and are listed in the table. 
\begin{table}
  \begin{center}
    \begin{tabular}{l|c|c|c}  
      \textbf{Intermediate} & \textbf{total no.} & \textbf{(+ve) terms} &\textbf{(-ve) terms}\\
      \textbf{energies} & \textbf{of terms} &  &  \\
      \hline
      4 4 4 4 4 & 96 & 48 & 48\\
      4 8 8 8 4 & 192 & 96 & 96\\
      4 4 8 8 4 & 96 & 48 & 48\\
      4 8 4 8 4 & 24 & 0 & 24\\
      4 8 8 4 4 & 96 & 48 & 48\\
      4 4 4 8 4 & 48 & 24 & 24\\
      4 4 8 4 4 & 48 & 24 & 24\\
      4 8 4 4 4 & 48 & 24 & 24\\
      4 8 12 8 4 & 72 & 0 & 72\\
    \end{tabular}
    \caption{Different terms contributing to $B_{p}^{(6)}$} \label{tab:table1}
  \end{center}
\end{table}
Observe that only the (4,8,4,8,4) and (4,8,12,8,4) processes give non-zero contributions, and the rest of them cancel with each other. We finally obtain
\begin{align}
H_{\text{eff}}^{(6)}=+\frac{3 J_{x}^{3}J_{y}^{3}}{256 J_{z}^{5}}B_{p}^{(6)}\ \ \ \text{where}\ B_{p}^{(6)}=\prod_{j=1}^{6}\tau^{z}_{j}
\end{align}
Tenth order: The next non-trivial correction comes at this order of expansion, which generates $B_{p}^{(10)}$. There are total 10! possible ways of acting the perturbations and we numerically determine the corresponding coupling constant,
\begin{align}
H_{\text{eff}}^{(10)}= + \frac{35}{2^{16}}\frac{J_{x}^{5}J_{y}^{5}}{J_{z}^{9}}B_{p}^{(10)},\ \ \ \text{where }\ B_{p}^{(10)}=\prod_{j=1}^{10}\tau^{z}_{j}
\end{align}
Here also, we neglect the ``disconnected plaquette'' contributions, for example $\sim (A^{(m)}_{v})_{i}(B^{(n)}_{p})_{j}$, with $m+n<10$, since these terms only trivially renormalize the dominant lower order coupling constants. \\

Apparently, the signs of the different coupling parameters in the TC Hamiltonian imply a ground state flux configuration that is different from what has been found numerically in Ref.\cite{PhysRevB.110.214438}. This apparent discrepancy resolves once the flux operators of $H_{\text{eff}}$ (i.e., $A_{v}^{(n)}$ and $B_{p}^{(n)}$) are correctly transformed back to those expressed in terms of the original site-spins ($\sigma^{\alpha}$), i.e., $W_{p}^{(n)}(\lbrace \alpha\rbrace)=\prod_{\langle ij\rangle \partial P}\sigma^{\alpha_{ij}}_{i}\sigma^{\alpha_{ij}}_{j}$, where $n=$ number of sites in a plaquette, $p$ of the tri-coordianted lattice. These two representations (of the flux operators) are related by an overall minus sign in some cases, as shown below,
\begin{align}
&(1)\ \ W^{(4)}_{p}(x,z)=(-1)\prod_{r=1}^{4}\sigma^{y}_{r}=(-1)\prod_{j=1}^{2}\tau^{x}_{j}=-A_{v}^{(2)} \ \ \ 
&(2)\ \ W^{(4)}_{p}(x,y)=(-1)\prod_{r=1}^{4}\sigma^{z}_{r}=(-1)\prod_{j=1}^{4}\tau^{z}_{r}=-B_{p}^{(4)}\nonumber \\
&(3)\ \ W^{(6)}_{p}(x,z)=(-1)\prod_{r=1}^{6}\sigma^{y}_{r}=(+1)\prod_{j=1}^{3}\tau^{x}_{j}=A_{v}^{(3)}\ \ \ 
&(4)\ \ W^{(6)}_{p}(y,z)=(-1)\prod_{r=1}^{6}\sigma^{x}_{r}=(-1)\prod_{j=1}^{3}\tau^{x}_{r}=-A_{v}^{(3)}\nonumber\\
&(5)\ \ W^{(6)}_{p}(x,y)=(-1)\prod_{r=1}^{6}\sigma^{z}_{r}=(-1)\prod_{j=1}^{6}\tau^{z}_{r}=-B_{p}^{(6)}\ \ \ 
&(6)\ \ W^{(8)}_{p}(y,z)=(-1)\prod_{r=1}^{8}\sigma^{x}_{r}=(-1)\prod_{j=1}^{4}\tau^{x}_{r}=-A_{v}^{(4)}\nonumber\\
&(7)\ \ W^{(10)}_{p}(y,z)=(-1)\prod_{r=1}^{10}\sigma^{x}_{r}=(-1)\prod_{j=1}^{5}\tau^{x}_{r}=-A_{v}^{(5)}\ \ \ 
&(8)\ \ W^{(10)}_{p}(x,z)=(-1)\prod_{r=1}^{10}\sigma^{y}_{r}=(+1)\prod_{j=1}^{5}\tau^{x}_{r}=A_{v}^{(5)}\nonumber\\
&(9)\ \ W^{(10)}_{p}(x,y)=(-1)\prod_{r=1}^{10}\sigma^{z}_{r}=(-1)\prod_{j=1}^{5}\tau^{z}_{r}=-B_{p}^{(10)}
\end{align}

Therefore, we adopt a sign convention for the toric-code Hamiltonian ($H_{\text{eff}}$) that reproduces the same ground-state flux configurations as those obtained in the original site-spin representation. This allows us to directly relate the results derived for $H_{\text{eff}}$ to those of the parent Kitaev model in Ref.\cite{PhysRevB.110.214438}. The final list of the coupling parameters used in the main text is given in the following table \ref{tab:table4}. 
\begin{table}
  \begin{center}
    \renewcommand{\arraystretch}{1.5} 
    \begin{tabular}{c|c|c} 
      $n$ & $\lambda_A^{(n)}$ & $\lambda_B^{(n)}$\\
      \hline 
      $2$ &  $+ (J^{2}/2J_{z})$ & $0$\\ 
      \hline 
      $3$ &  $-(3J^{3}/8J_{z}^{2})$  & $0$\\ 
      \hline
      $4$ & $+(5J^{4}/16J_{z}^{3})$ & $+(J^{4}/16J_{z}^{3})$\\
      \hline
      $5$ & $-(35 J^{5}/128 J_{z}^{4})$ & $0$\\
      \hline
      $6$ & $0$ & $-(3J^{6}/2^{8}J_{z}^{5})$\\
      \hline
      $10$ & $0$ & $-(35J^{10}/2^{16}J_{z}^{9})$
    \end{tabular}
    \caption{The list of $A_{v}^{(n)}$ and $B^{(n)}_{p}$ coupling parameters of the toric code used in the main text. We consider the case: $J_{x}=J_{y}=J\ (\ll J_{z})$. See Fig.\ref{fig:TC-coupling} for the distribution of these couplings for a specific generation.} 
    \label{tab:table4}
  \end{center}
\end{table}

\begin{figure}
    \centering
    \includegraphics[width=0.9\linewidth]{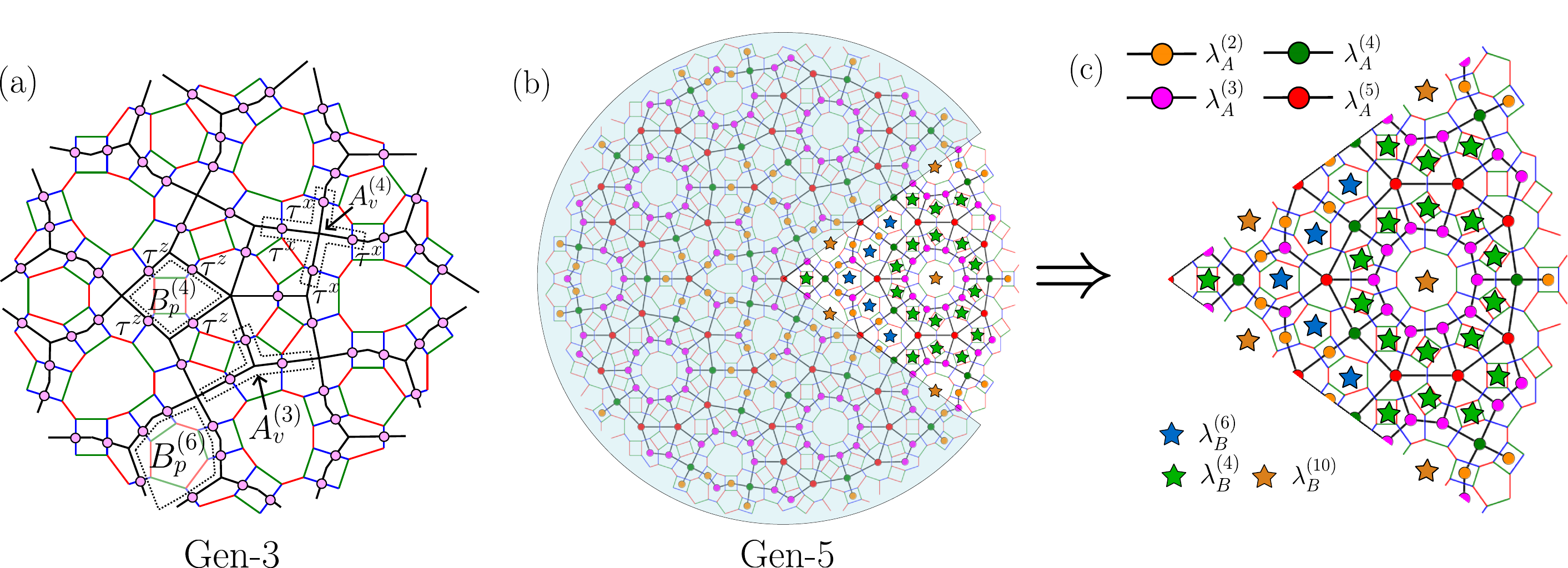}
    \caption{{\bf Terms of the toric code Hamiltonian}: (a) Different types of star ($A_{v}^{(n)}$) and plaquette ($B_{p}^{(n)}$) operators are shown for Gen-3 lattice. The distribution of coupling parameters ($\lambda_{A}^{(n)}$, $\lambda_{B}^{(n)}$) is demonstrated for larger Gen-5 lattice in Fig.(b) and (c).} 
    \label{fig:TC-coupling}
\end{figure}

\section{Toric code Hamiltonian for a different strong-bond anisotropy}
\begin{figure}
    \centering
    \includegraphics[width=0.5\linewidth]{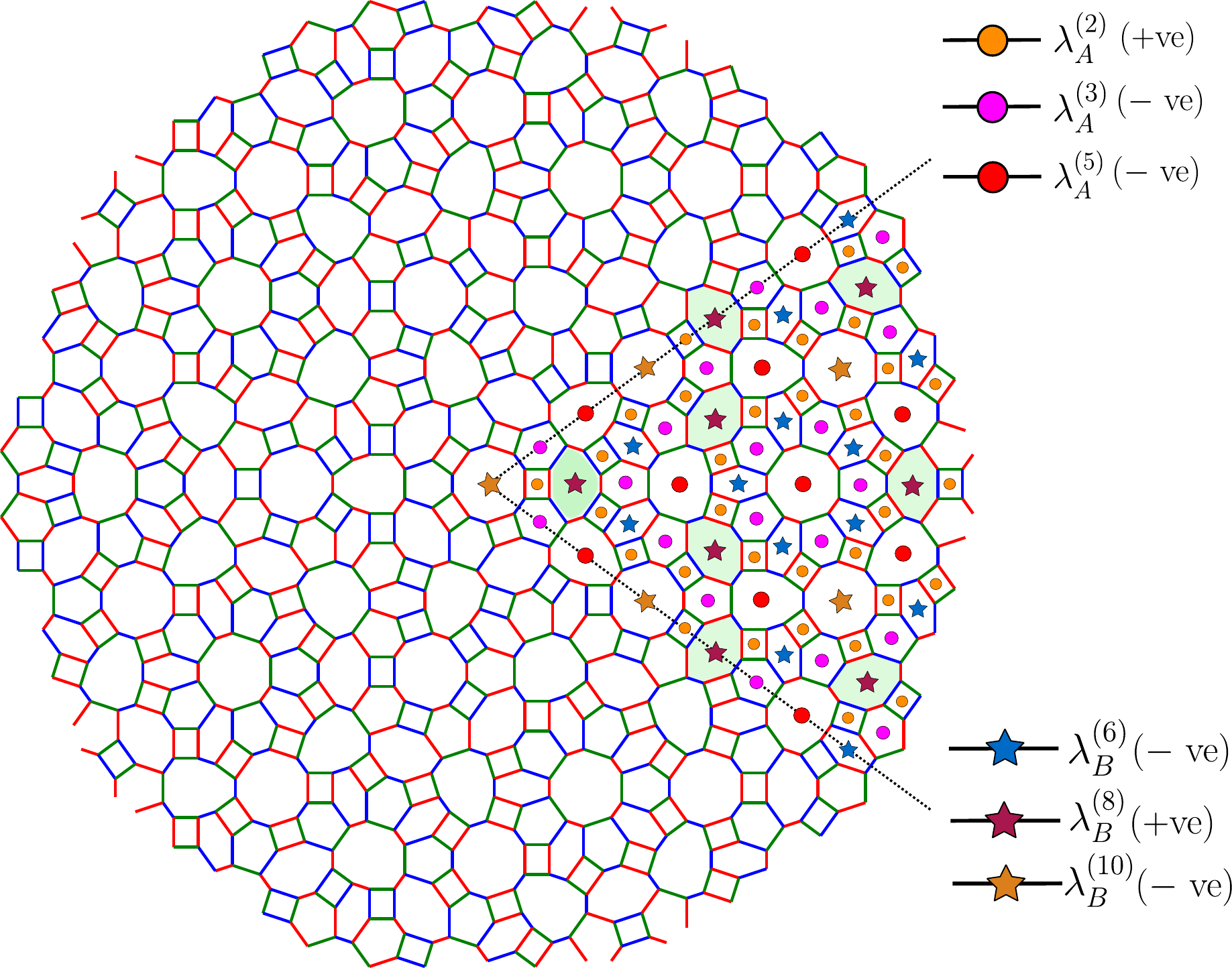}
    \caption{{\bf Toric code Hamiltonian in the $J_{x}\gg J_{y}, J_{z}$ limit:} Shown are the different types of star ($A_{v}^{(n)}$) and plaquette ($B_{p}^{(n)}$) operators, indicated by colored circles and star symbols, respectively, for the Gen-5 lattice within an angular cross-section. The ground-state background flux configuration is represented by the green shaded regions.}
    \label{fig:TC-diff-anisotropy}
\end{figure}

Instead of taking the large-$J_{z}$ limit, if one chooses a different coupling to dominate, the resulting toric-code Hamiltonian acquires a modified structure. Both the geometry of the effective toric-code lattice—on which the $e$ anyons reside on the vertices and the $m$ anyons at the plaquette centers—and the relative densities of the various $A_{v}$ and $B_{p}$ operators are altered. In addition, the corresponding coupling constants associated with these operators change accordingly. For illustration, Fig.~\ref{fig:TC-diff-anisotropy} shows the distribution of star ($A_{v}$) and plaquette ($B_{p}$) terms along with their respective coupling strengths in the $J_{x}\gg J_{y}, J_{z}$ limit. By comparing with the large-$J_{z}$ case (Fig.\ref{fig:TC-coupling}), we see that the roles of the star and plaquettes operators have flipped in some cases. The background $\pi$-flux distribution of the zero-field ground state is also different (Compare with Fig.1(b) in the main text).

\section{Energy spectra for spatially non-uniform and uniform $\lambda_{A}^{(n)}$}
\begin{figure}[h!]
    \centering
    \includegraphics[width=0.9\linewidth]{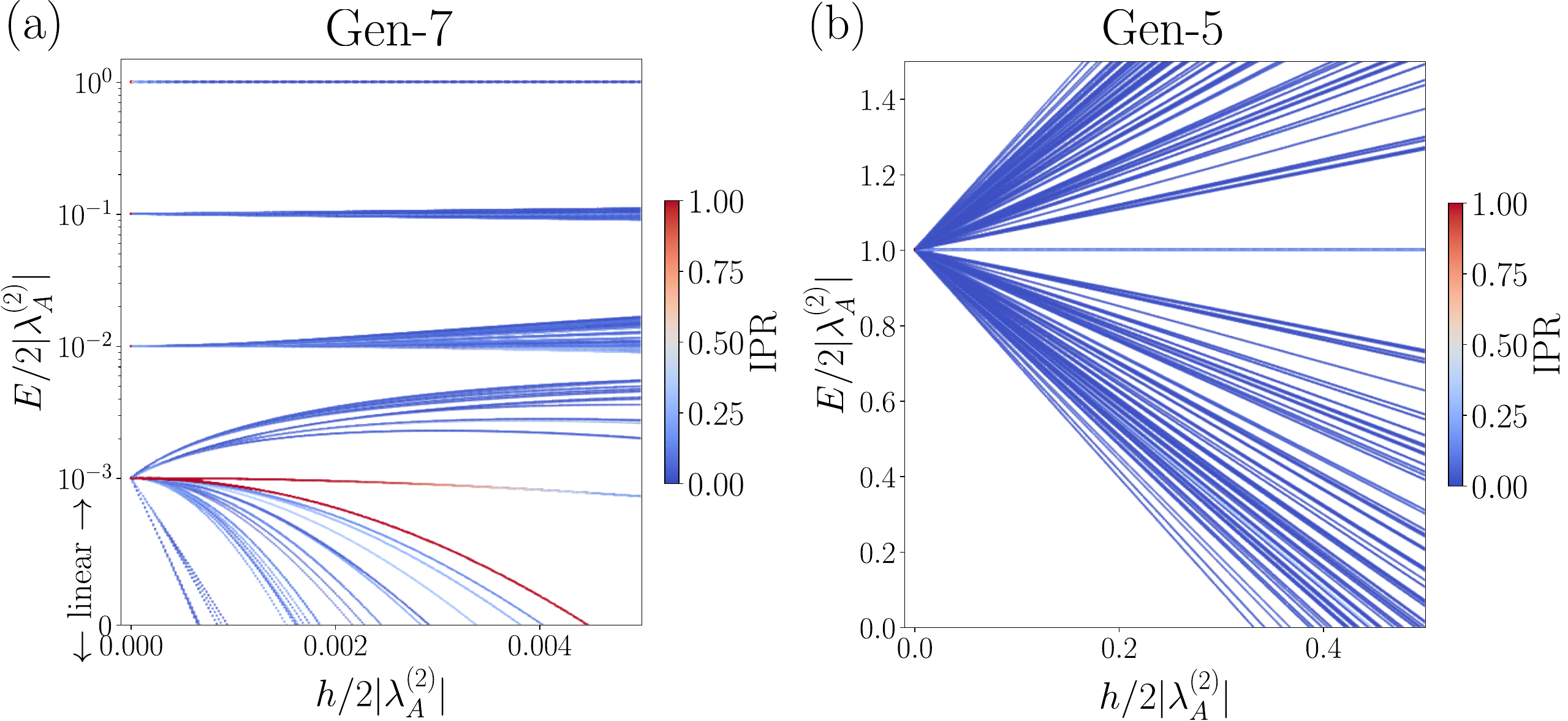}
    \caption{{\bf The $e$ anyon energy spectra}: (a) Energy eigenvalues of the effective low-energy bosonic hopping model [Eq.(14) of main text] with non-uniform onsite potentials $\lambda_{A}^{(n)}$ are plotted as a function of the Zeeman coupling $h$ for the Gen-7 lattice (1906 sites). The characteristic four-band structure—exponentially separated in energy—and both linear ($\sim h$) and quadratic ($\sim h^{2}$) scaling of the levels, observed earlier for the Gen-5 lattice (see main text), persist here. (b) The spectrum in the absence of onsite potential non-uniformity ($\lambda_{A}^{(n)} = \lambda_{A}^{(2)}$ for all $n$) is shown as a function of $h$. In this case, all energy levels exhibit linear scaling ($\sim h$). The Gen-7 spectrum shows a similar behavior, but with a denser distribution of levels owing to the larger system size.} 
    \label{fig:spectrum-supplement}
\end{figure}

See Fig.\ref{fig:spectrum-supplement} for the low-energy single–$e$–anyon spectra in two representative cases: (1) the Gen-7 lattice with spatially non-uniform onsite potentials $\lambda_{A}^{(n)}$ (see Eq.(14) of main text), and (2) the Gen-5 lattice with uniform onsite potentials $\lambda_{A}^{(n)} = \lambda_{A}^{(2)}$ for all $n$, and in presence of the background $\pi$-flux distribution.

\section{Qualitative understanding of $E$ vs. $h$ scaling}

\begin{figure}
    \centering
    \includegraphics[width=0.4\linewidth]{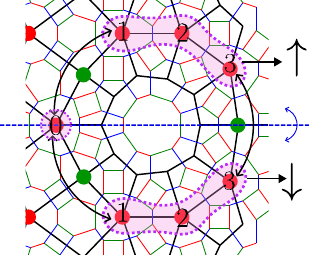}
    \caption{{\bf Effective ring model for low-lying $A_{v}^{(5)}$ eigenstates:} To obtain the low-energy branches (b1–b4) of the $A_{v}^{(5)}$ eigenstates, we consider an effective model in which the upper ($\uparrow$) and lower ($\downarrow$) three-site chains (red circles) are coupled to each other, as well as to an isolated site (0), via hopping processes mediated through high-energy intermediate sites (green).} 
    \label{fig:coupled-3-site}
\end{figure}

A qualitative insight into why the energy ($E$)–Zeeman field ($h$) scaling differs for various low-energy states (as shown in Fig.~(3(c)) of the main text) can be obtained by analyzing the low-lying eigenstates of the circular contour illustrated in Fig.~\ref{fig:coupled-3-site}. This contour consists of two three-site tight-binding chains, positioned symmetrically above and below a mirror reflection axis (indicated by the dashed blue line), and labeled by $\uparrow$ and $\downarrow$ orbitals, respectively. In addition, there is a single isolated site (labeled 0) that has the same onsite energy $V_{R}$ as that of the sites of the three-site chains. The upper and lower chains, as well as the lone site, are interconnected via hopping through high-energy ``bridge'' sites (green dots).

The 3-site nearest-neighbor hopping model can be diagonalized by hand, and we obtain the following three single-particle eigenstates,
\begin{align}
E_{1}=V_{R}-\sqrt{2}h\ \ \ \ ,& \ \ \ \ \ket{E_{1}}=\frac{1}{2}(c_{1}^{\dag}+\sqrt{2}c_{2}^{\dag}+c_{3}^{\dag})\ket{0}\nonumber\\
E_{2}=V_{R}\ \ \ \ ,& \ \ \ \ \ket{E_{2}}=\frac{1}{\sqrt{2}}(c_{1}^{\dag}-c_{3}^{\dag})\ket{0}\nonumber\\
E_{3}=V_{R}+\sqrt{2}h\ \ \ \ ,& \ \ \ \ \ket{E_{3}}=\frac{1}{2}(c_{1}^{\dag}-\sqrt{2}c_{2}^{\dag}+c_{3}^{\dag})\ket{0} \label{3-site-eigfns}
\end{align}
Here, $V_{R}=2|\lambda_{A}^{(5)}|$ is the onsite potential at the $A^{(5)}$ (red) sites, and $h$ is the hopping amplitude within the three-site units. 
The full microscopic model of the ring is symmetric under mirror reflection. Therefore, the energy eigenstates can be divided into $\mathbb{Z}_{2}$ reflection symmetric and anti-symmetric eigenstates. For the anti-symmetric eigenstates, there must be nodes along the reflection symmetry line. There are three such eigenstates, which are given by,
\begin{align}
\ket{\psi}_{\text{anti-symm}}\sim (\ket{E_{\alpha, \uparrow}}-\ket{E_{\alpha,\downarrow}})\ \ \ \ \alpha=1,2,3 \label{wavefn-anti-symm}
\end{align}
up to some normalization constant. The energies are the same as those of the isolated 3-site problem. Only when the ring is connected back to the 2D lattice, corrections of order $\sim O(h^{2}/2|\lambda_{A}^{(n)}|)$ (with $n<5$) will appear from virtual hopping processes out of the ring. The crucial point here is that the wavefunction amplitudes at the ``bridge'' sites and at site 0 is zero. This is what happens for branches b1 and b3 (Fig.\ref{fig:A5-eigenfns}) which respectively correspond to the cases $\alpha=1$ and $\alpha=2$ in Eq.\eqref{wavefn-anti-symm}. 

For the symmetric states, using the fact that the onsite potential at the ``bridge'' (green) sites, $V_{G}\sim |\lambda_{A}^{(4)}|\gg V_{R}\ , h$, we integrate out these high-energy sites and obtain the following 7-site low-energy model,
\begin{align}
\tilde{H}_{\text{ring}}&=(V_{R}-\frac{2h^{2}}{V_{G}}\big)b^{\dag}_{0}b_{0}+(V_{R}-\frac{h^{2}}{V_{G}}\big)\sum_{\sigma}\sum_{i=1,3}b^{\dag}_{i\sigma}b_{i\sigma}+ V_{R}\sum_{\sigma}b^{\dag}_{2\sigma}b_{2\sigma}\nonumber\\
&-h\sum_{i=1,2}(b^{\dag}_{i\sigma}b_{i+1,\sigma}+h.c.) - \frac{h^{2}}{V_{G}}(b^{\dag}_{3\uparrow}b_{3,\downarrow}+h.c.)-\frac{h^{2}}{V_{G}}(b^{\dag}_{1\uparrow}b_{0}+b^{\dag}_{1\downarrow}b_{0}+h.c.) \label{H-ring-approx}
\end{align}

\begin{figure}
    \centering
    \includegraphics[width=0.9\linewidth]{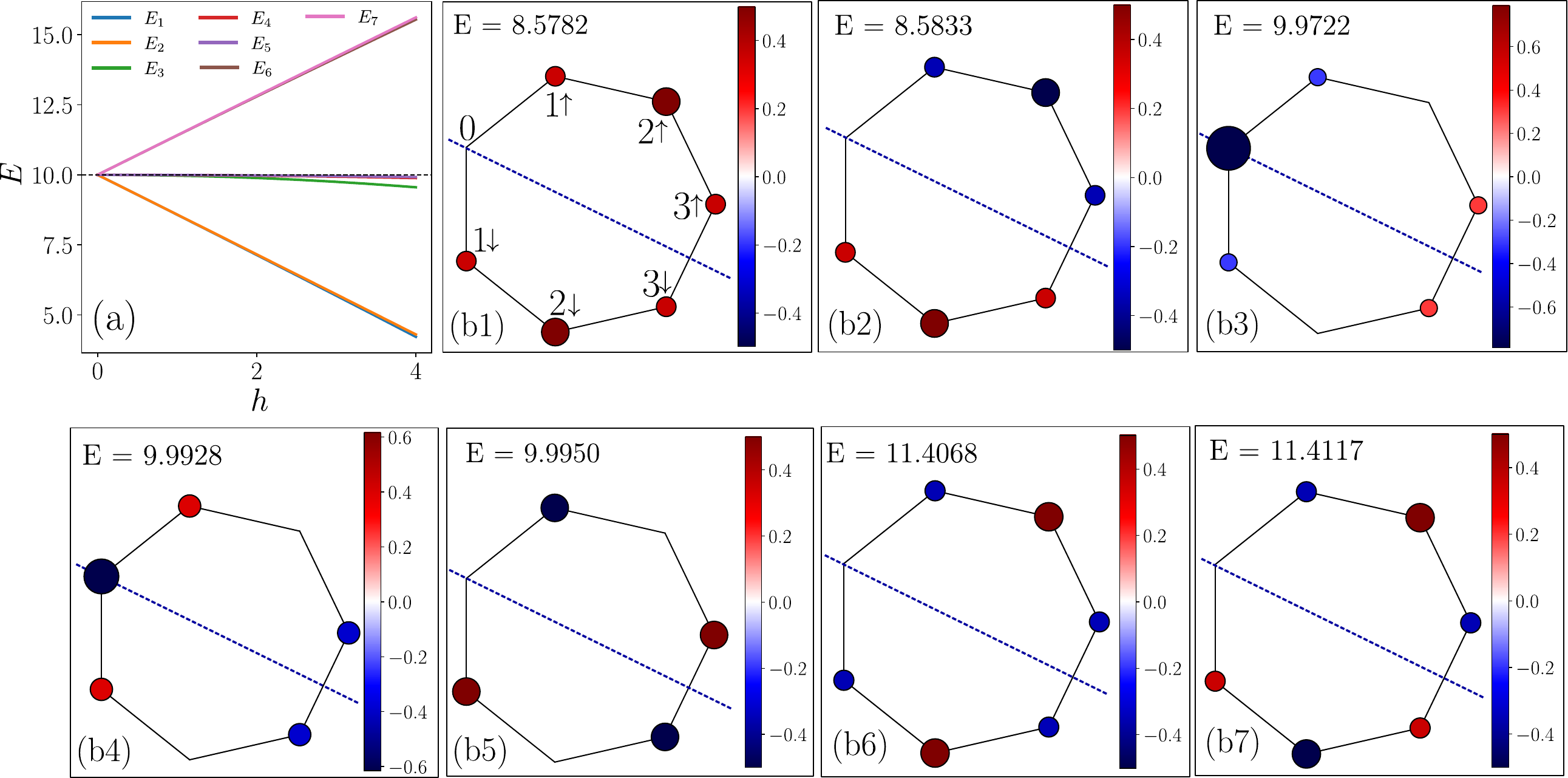}
    \caption{Eigenvalues and eigenfunctions of $\tilde{H}_{\text{ring}}$ (Eq.~\eqref{H-ring-approx}): The energy eigenvalues $E$ as a function of the hopping amplitude $h$ are shown in panel (a) for $V_{R}=10.0$ and $V_{G}=100.0$. Panels (b1)–(b7) display the corresponding energy eigenfunctions for the same values of $V_{R}$ and $V_{G}$ at fixed $h=1.0$. The conventions for labeling the lattice sites are shown only in panel (b1).}
    \label{fig:ring-model-spectrum}
\end{figure}

We diagonalize the effective Hamiltonian (Eq.~\eqref{H-ring-approx}) and obtain its eigenspectra, shown in Fig.~\ref{fig:ring-model-spectrum}. The onsite potentials are fixed at $V_{R} = 10.0$ and $V_{G} = 100.0$. For $h=1.0$, the spectrum contains both symmetric ($E = 8.5782, 9.9722, 9.9928, 11.4068$) and antisymmetric ($E = 8.5833, 9.9950, 11.4117$) eigenstates. As discussed earlier, the antisymmetric states exhibit nodes at the $0$-th site.

The lowest two states, $E_{1} = 8.5782$ and $E_{2} = 8.5833$ (for $h=1.0$), follow the dispersion relation $E \sim V_{R} - \sqrt{2}h$, with small quadratic corrections ($\sim h^{2}$) appearing only for the symmetric combination of $\uparrow$ and $\downarrow$ units. These states correspond to the branch b1 in Fig.~\ref{fig:A5-eigenfns} of the full quasicrystal lattice.

There are three states, $E_{3} = 9.9722$, $E_{4} = 9.9928$, and $E_{5} = 9.9950$, that lie nearly parallel to the line $E = V_{R} = 10.0$. Among them, the states $E_{3}$ and $E_{4}$ exhibit large amplitudes at site $0$; these originate from the hybridization between the $0$th orbital and the symmetric linear combination of the three-site eigenstates $\ket{E_{2}\sigma}$ (defined in Eq.\eqref{3-site-eigfns}). These states correspond to branches b2 and b4 in Fig.\ref{fig:A5-eigenfns}, and their energies scale as $E \sim V_{R} - O(h^{2}/V_{G})$.

The ring eigenstate $E_{6}$ $\sim \ket{E_{2\uparrow}} - \ket{E_{2\downarrow}}$, which has a vanishing amplitude at the high-energy sites due to its antisymmetric nature. Consequently, it does not acquire any $h^{2}/V_{G}$ correction. However, an $O(h^{2})$ correction may appear when this ring model is embedded within the full two-dimensional quasicrystalline lattice. This eigenstate corresponds to branch b3 in Fig.~\ref{fig:A5-eigenfns}.

\begin{figure}
    \centering
    \includegraphics[width=0.6\linewidth]{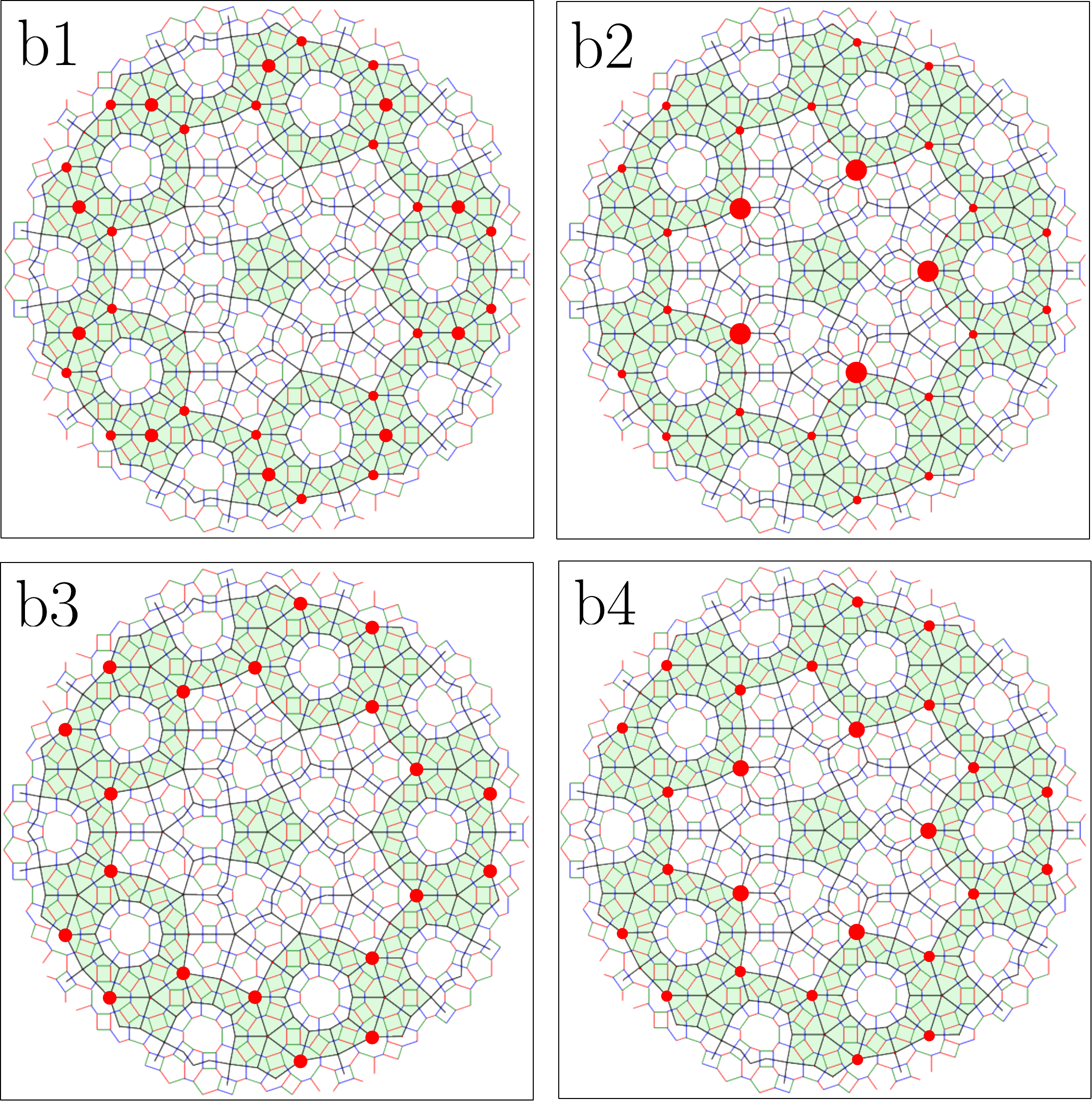}
    \caption{The probability density ($|\psi(r)|^{2}$) of energy eigenstates belonging to the four branches (b1 to b4) shown in the Main text Fig.(3) (by red dashed curves), for the Zeeman coupling, $h/2|\lambda_{A}^{(2)}|=0.0005$.  } 
    \label{fig:A5-eigenfns}
\end{figure}

\section{\textcolor{blue}{Scaling of plateau residence time and analogy with ring-hopping model}}

\textcolor{blue}{In this section, we present supplementary results regarding the scaling of plateau residence time ($T_{p}$) as a function of $g = |\lambda_{A}^{(n+1)}|/|\lambda_{A}^{(n)}|$, for two different initial conditions ($A_{v}^{(3)}$ sites) and several values of $h/|\lambda^{(2)}_A|$.}

\begin{figure}
    \centering
    \includegraphics[width=1.0\linewidth]{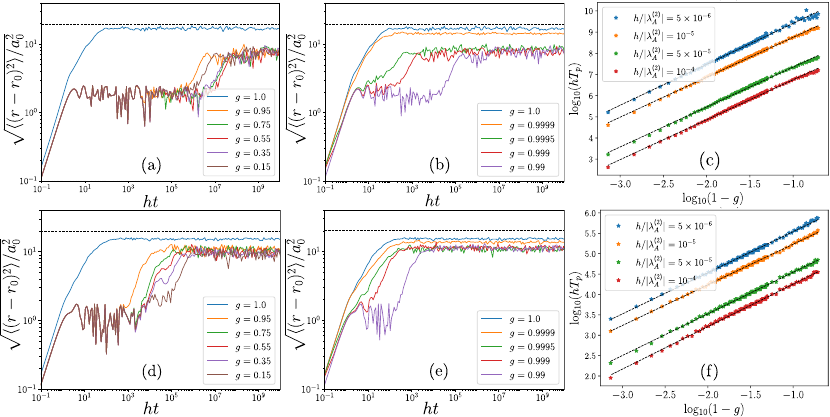}
    \caption{(a) and (d): Time-variation of mean-square displacement, $d(r_{0},t)$ for a wavepacket initialized at two different $A^{(3)}_{v}$-type vertex (same as those chosen in Fig.6(a) and 6(b) of the main text), for different values of the onsite coupling anisotropy, $g=|\lambda_{A}^{(n+1)}|/|\lambda_{A}^{(n)}|$. (b) and (c): The same function is plotted for $g$ values close to 1, to demonstrate how rapidly the plateau width decreases with $g$, near $g=1$. (c) and (f): Scaling of plateau residence time ($T_{p}$) as a function of $1-g$ shows power-law decrease in $T_{p}$ near $g=1$, with exponent $\alpha_p\approx 1.9$ (Fig.6(a), main text) and $\alpha_p\approx 1.0$ (Fig.6(b), main text), which are roughly independent of $h/|\lambda^{(2)}_A|$. These results are for the Gen-7 lattice.} 
    \label{fig:plateau-time}
\end{figure}

\textcolor{blue}{As in the main text, the wavepackets are initially localized on two different $A^{(3)}$-type vertices belonging to the class of sites connected through $O(h)$ hopping processes. The corresponding time-evolved probability distributions are presented in Fig.~6(a) and (b) of the main text. In Fig.\ref{fig:plateau-time}(a) and (d), we plot the mean-square displacement $d(r_0,t)$ as a function of time for several values of $g$. These results explicitly demonstrate that, in the strong-coupling regime (small $g$), the evolution of the plateau width depends sensitively on the choice of the initial state. For the case shown in Fig.\ref{fig:plateau-time}(d), the plateau width decreases monotonically with increasing $g$, consistent with the behavior reported in Fig. 6(d) of the main text. In contrast, the dynamics shown in Fig.\ref{fig:plateau-time}(a) exhibit a non-monotonic trend: the plateau width initially increases as $g$ is raised from $0.15$ to $0.55$, thereby corroborating the behavior observed in Fig. 6(c) of the main text.}

\textcolor{blue}{In Fig.\ref{fig:plateau-time}(c) and (f), we show the scaling of $T_p$ as a function of $1-g$ in the log-scale, near $g=1$ (boxed areas in Fig.6(c) and (d) of the main text). As clearly visible from these figures, the scaling exponent has almost no dependence on the perturbative Zeeman coupling $h/|\lambda_A^{(2)}|$.}

\begin{figure}
    \centering
    \includegraphics[width=1.0\linewidth]{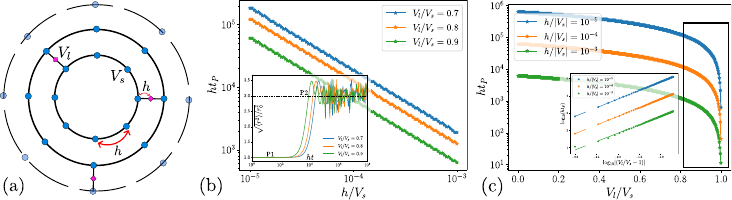}
    \caption{(a) A model of coupled rings: each ring consists of a nearest-neighbor hopping model with hopping amplitude $h$ and an onsite potential $V_{s}$ (identical for all rings). The rings are coupled via hopping through a \textit{bridge site} (purple diamond), which has a different onsite potential $V_{l}$. (b) Plateau residence time ($t_P$) as a function of $h/V_{s}$, for different $V_{l}/V_{s}$. The wavepacket spreading distance $d(t)=\sqrt{\bra{\psi_0(t)}r^2\ket{\psi_0(t)}}$ as a function of time ($ht$) depicts two distinct plateaus (P1, P2) for different ratios $V_l/V_s$ and fixed effective hopping strength $h/|V_s|=10^{-4}$ (Fig.(b) inset). (c) Plateau residence time ($t_P$) vs. $V_l/V_s$, for several $h/V_s$ values. Scaling of $t_P$ near $V_l/V_s=1$ (boxed region) is shown in the inset. \label{fig:plateau-time-ring-model}}
\end{figure}

\textcolor{blue}{To demonstrate the non-triviality of wavepacket dynamics in the quasicrystalline toric code, we compare it to the wavepacket dynamics in a simple toy model of periodically connected rings, shown in Fig.\ref{fig:plateau-time-ring-model}(a).} The model is motivated by the geometrical structure of the contours along which wavepacket delocalization occurs in Fig.~6(a) of the main text. We consider a system of concentric rings, each consisting of the same number of sites and described by a nearest-neighbor hopping Hamiltonian with hopping amplitude $h$ and onsite potential $V_{s}$ (identical for all rings). Only one site of each ring couples to a single site of the adjacent ring, with the coupling mediated through a \textit{bridge site} at onsite potential $V_{l}$, distinct from $V_{s}$. The hopping amplitude between a ring site and the bridge site is taken to be the same as the intra-ring hopping ($h$). Each ring mimics the smaller contours in Fig.~6(a) composed exclusively of $A^{(3)}$-type vertices, whereas the bridge site mimics the combined effect of the intermediate $A^{(5)}$ and $A^{(4)}$ vertices, whose onsite potentials are lower than those of the $A^{(3)}$ sites. Thus, in our context, $V_{l} < V_{s}$.

\textcolor{blue}{In Fig.~\ref{fig:plateau-time-ring-model}(b) and (c), we show the plateau transition time $t_{P}$ as a function of the hopping strength $h/V_s$ and the barrier strength $V_l/V_s$ for a system comprising 6 concentric rings, with each ring containing 20 sites. The inter-ring hopping links are arranged such that no two consecutive inter-ring bonds share a common ring site. The outermost ring is subsequently connected back to the innermost ring through bridge-site-mediated hopping.}

\textcolor{blue}{For this geometry, the time evolution of the wavepacket exhibits two well-defined plateaus: the first plateau (P1) occurs at $d(t)=\sqrt{\langle r^2\rangle_t}=1$, while the second (P2) appears at $d(t)\approx 3$, measured in units of the inter-ring distance (see the inset of Fig.~\ref{fig:plateau-time-ring-model}(b)). The distance here is measured from the origin ($r_0=0$), in contrast to Fig.~6 of the main text where the distance is measured with respect to the initial location of the delta function wavepacket. Although the qualitative dependence of the plateau transition time on the model parameters is similar to that observed in the quasicrystal case (Fig.~6, main text), namely, the power-law behavior, the exponents obtained for the ring model agree with the predictions of perturbation theory. This agreement is in contrast to the quasicrystal case, where the corresponding exponents deviate from perturbative expectations. Specifically, we find here $t_{P}^{-1}\sim \frac{h^2}{\Delta V},$
where $\Delta V=|V_S-V_l|$. Thus, the plateau transition time $t_P$ is essentially determined by the effective hopping scale between consecutive rings.}

\section{Wavepacket evolution for various $A^{(5)}$ and $A^{(3)}$ $O(h^n)$ vertices for $n>1$}

\begin{figure}
    \centering
    \includegraphics[width=0.6\linewidth]{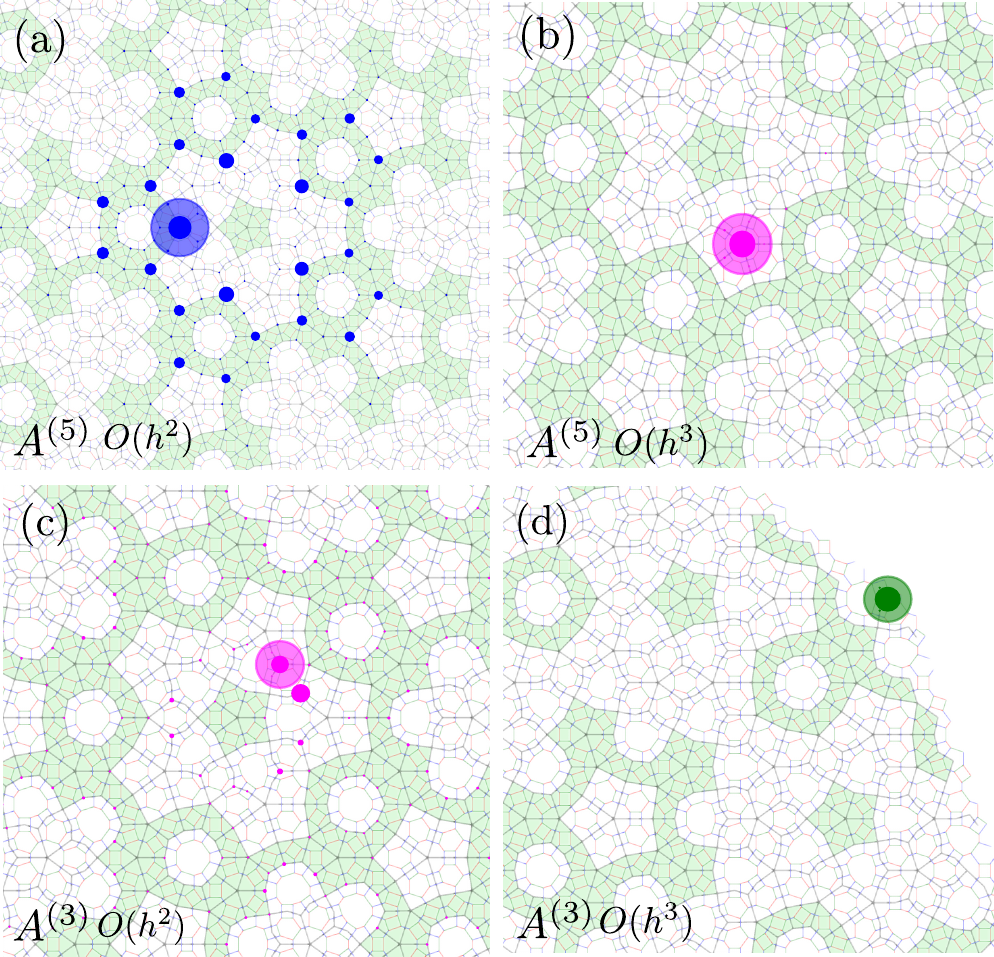}
    \caption{Time averaged probability density of $e$ anyons for different initial conditions: (a) $A^{(5)}$, $O(h^2)$ vertex, (b) $A^{(5)}$, $O(h^3)$ vertex, (c) $A^{(3)}$, $O(h^2)$ vertex, and (d) $A^{(3)}$, $O(h^3)$ vertex.    }
    \label{fig:wavepacket-dynamics-supp-ngtr1}
\end{figure}

\textcolor{blue}{In Fig.~\ref{fig:wavepacket-dynamics-supp-ngtr1}, we present supplementary wavepacket-dynamics results corresponding to Fig.~5 of the main text. Specifically, we examine the spreading of $e$-anyon wavepackets initialized at $A^{(5)}$ and $A^{(3)}$ vertices that belong to the class of sites connected through higher-order hopping processes, $O(h^n)$ with $n>1$. The observed dynamics are fully consistent with the behavior shown in the mean-square displacement $d(r_0,t)$ plots in Fig.~5(a) and (c) of the main text. There are no bulk sites belonging to the $A^{(3)}$, $O(h^3)$ group; therefore, in this case, we show the wavepacket evolution for a boundary vertex. }

\section{Derivation of $\pi$-flux localized eigenstate}
\begin{figure}[h!]
    \centering
    \includegraphics[width=0.5\linewidth]{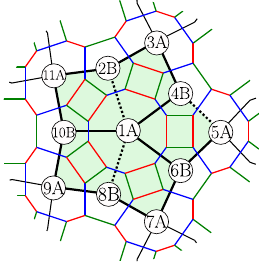}
    \caption{A nearest-neighbor hopping model on a five-cornered \text{star}-like subsystem containing odd number of $\pi$ fluxes (green shaded plaquettes). To thread these fluxes, we change the sign of the hopping on particular bonds (dashed lines).} 
    \label{fig:star-cluster}
\end{figure}

In this section, we analyze how a localized $e$-anyon eigenstate can emerge within a star-shaped cluster (Fig.~\ref{fig:star-cluster}) of the quasicrystalline toric code lattice that encloses an odd number of $\pi$ fluxes. The cluster contains 11 sites (vertices), which we decompose into two sublattices, $A$ and $B$. Importantly, this cluster is coupled to the rest of the 2D lattice only through the bonds emanating from its five sharp corner sites labeled $(2n+1)A$ with $n=1,\dots,5$. A necessary condition for localization within the cluster is that the wavefunction develops nodes at these corner sites; such nodal structure prevents tunneling out of the cluster and thus inhibits leakage into the surrounding lattice.

To identify such localized states, we analytically solve the low-energy bosonic hopping Hamiltonian (Eq.~(13) in the main text) defined on the 11-site cluster, assuming uniform onsite potentials $|\lambda_{v}^{(n)}|=\lambda$ across all sites. Our goal is to search for eigenstates whose amplitudes vanish precisely at the $(2n+1)A$ vertices, thereby yielding confined, cluster-localized solutions.

The Hamiltonian matrix for this 11-site cluster is given by,
\begin{align}
H = \ \ \ 
\begin{array}{c|cccccc|cccccc}
      & 1A & 3A & 5A & 7A & 9A & 11A & 2B & 4B & 6B & 8B & 10B \\ \hline
1A    & \lambda & 0 &  0 &  0 &  0 & 0  &  +h & -h &  -h &  +h &  -h \\
3A    & 0 & \lambda & 0 &  0 &  0 &  0  & -h &  -h & 0 &  0 &  0 \\
5A    &  0 & 0 & \lambda & 0 &  0 &  0  &  0 & +h &  -h & 0 &  0  \\
7A    &  0 &  0 & 0 & \lambda & 0 &  0  &  0 &  0 & -h &  -h & 0   \\
9A    &  0 &  0 &  0 & 0 & \lambda & 0  &  0 &  0 &  0 & -h &  -h  \\
11A   &  0 &  0 &  0 &  0 & 0 & \lambda  & -h &  0 &  0 &  0 & -h  \\ \hline
2B    &  +h & -h &  0 &  0 &  0 & -h  & \lambda & 0 &  0 &  0 &  0 \\
4B    & -h &  -h & +h &  0 &  0 &  0  & 0 & \lambda & 0 &  0 &  0 \\
6B    &  -h & 0 &  -h & -h &  0 &  0  &  0 & 0 & \lambda & 0 &  0  \\
8B    &  +h &  0 &  0 & -h & -h &  0  &  0 &  0 & 0 & \lambda & 0 \\
10B   &  -h &  0 &  0 & 0 &  -h & -h  &  0 &  0 &  0 & 0 & \lambda \\
\end{array}
\end{align}
We are interested in a finite energy solution of the form, 
\begin{align}
\psi = (
\psi_{1}, 0,0,0,0,0,\psi_{2},\psi_{4},\psi_{6},\psi_{8},\psi_{10})^{T}
\end{align}
The set of independent equations we obtain are
\begin{align}
&\psi_{2} - \psi_{4} -\psi_{6} +\psi_{8} -\psi_{10} = \bigg[\frac{\epsilon(\lambda,h)-\lambda}{h}\bigg]\psi_{1}\\
&\psi_{4}=-\psi_{2}\ \ ,\ \ \psi_{6}=\psi_{4}\ \ ,\ \ \psi_{8} =-\psi_{6}\ ,\ \psi_{10}=-\psi_{8}
\end{align}
where $\epsilon(\lambda,h)$ is the eigenenergy of the above state.
The solution we obtain is given by
\begin{align}
&\psi_{2}=\psi_{8}=\bigg[\frac{\epsilon(\lambda,h)-\lambda}{5h}\bigg]\psi_{1}\\
&\psi_{4}=\psi_{6}=\psi_{10}=-\bigg[\frac{\epsilon(\lambda,h)-\lambda}{5h}\bigg]\psi_{1}
\end{align}
where $\psi_{1}$ is fixed by wavefunction normalization.


\end{document}